\documentclass[pdflatex,sn-mathphys-num]{sn-jnl}

\usepackage{graphicx}%

\usepackage{amsmath,amssymb,amsfonts}%
\usepackage{multirow}%
\usepackage{amsthm}%
\usepackage{mathrsfs}%

\usepackage[title]{appendix}%
\usepackage[table]{xcolor}
\definecolor{mycolor}{RGB}{224,224,224}
\usepackage{textcomp}%
\usepackage{manyfoot}%
\usepackage{booktabs}%
\usepackage{algorithm}%
\usepackage{algorithmicx}%
\usepackage{algpseudocode}%
\usepackage{listings}%

\usepackage{lipsum} 
\usepackage{hyperref}
\usepackage{setspace}
\usepackage{rotating}
\usepackage{adjustbox}
\usepackage{subcaption}
\usepackage{array}
\usepackage{arydshln}

\theoremstyle{thmstyleone}%
\theoremstyle{thmstyletwo}%

\theoremstyle{thmstylethree}%

\begin{document}

\title{Towards accurate and reliable ICU outcome prediction: a multimodal learning framework based on belief function theory using structured EHRs and free-text notes}


\author[1,2]{\fnm{Yucheng} \sur{Ruan}}\email{yuchengruan@u.nus.edu}

\author[2]{\fnm{Daniel J.} \sur{Tan}}\email{djtan@u.nus.edu}

\author[2]{\fnm{See-Kiong} \sur{Ng}}\email{seekiong@nus.edu.sg}

\author*[1]{\fnm{Ling} \sur{Huang}}\email{iweisskohl@gmail.com}

\author[1,2]{\fnm{Mengling} \sur{Feng}}\email{ephfm@nus.edu.sg}

\affil[1]{\orgdiv{Saw Swee Hock School of Public Health}, \orgname{National University of Singapore}, \orgaddress{\country{Singapore}}}

\affil[2]{\orgdiv{Institute of Data Science}, \orgname{National University of Singapore}, \orgaddress{\country{Singapore}}}


\abstract{

Accurate Intensive Care Unit (ICU) outcome prediction is critical for improving patient treatment quality and ICU resource allocation. Existing research mainly focuses on structured data, e.g. demographics and vital signs, and lacks effective frameworks to integrate clinical notes from heterogeneous electronic health records (EHRs). 
This study aims to explore a multimodal framework based on belief function theory that can effectively fuse heterogeneous structured EHRs and free-text notes for accurate and reliable ICU outcome prediction. The fusion strategy accounts for prediction uncertainty within each modality and conflicts between multimodal data.
The experiments on MIMIC-III dataset show that our framework provides more accurate and reliable predictions than existing approaches.
Specifically, it outperformed the best baseline by 1.05\%/1.02\% in BACC, 9.74\%/6.04\% in F1 score, 1.28\%/0.9\% in AUROC, and 6.21\%/2.68\% in AUPRC for predicting mortality and PLOS, respectively. Additionally, it improved the reliability of the predictions with a 26.8\%/15.1\% reduction in the Brier score and a 25.0\%/13.3\% reduction in negative log-likelihood.
By effectively reducing false positives, the model can aid in better allocation of medical resources in the ICU. Furthermore, the proposed method is very versatile and can be extended to analyzing multimodal EHRs for other clinical tasks. The code implementation is available on \url{https://github.com/yuchengruan/evid_multimodal_ehr}.

}




\maketitle

\section{Introduction}\label{sec1}

The Intensive Care Unit (ICU) is a specialized hospital ward that offers comprehensive and continuous care to critically ill patients. As the population of critically ill patients grows, the demand for ICUs has risen significantly, placing strain on already limited and costly intensive care resources \cite{terwiesch2011working}, especially during public health crises like the COVID-19 pandemic, when hospitals face an overwhelming surge of patients \cite{arabi2021covid}.

Due to the limited availability of intensive care resources, researchers have emphasized the necessity of predicting ICU outcomes such as mortality rates and prolonged lengths of stay (PLOS). Accurate predictions can help in the efficient allocation of medical resources for patients in need and reduce unnecessary expenses without compromising patient care. Furthermore, they are crucial for healthcare providers in making informed decisions about patient care strategies and providing early interventions to patients at high risk of adverse outcomes \cite{halpern2010critical, wang2024timely}.

Over the past two decades, the adoption of electronic ICU technology has enabled the collection of extensive data on ICU patients, creating new opportunities for developing advanced methods to predict ICU outcomes.
Most previous research has concentrated on modeling ICU outcome predictions using structured EHR data \cite{iwase2022prediction, ashrafi2024deep, gao2024prediction}, which often captures only a portion of clinical information. It may miss out on the rich contextual information that unstructured EHR data (such as nursing notes, patient narratives, and imaging reports) can provide. Natural language processing (NLP) techniques have been well explored to extract valuable insights from unstructured free-text EHR data \cite{sheikhalishahi2019natural, koleck2021identifying, oliwa2021development, zhou2023extracting}.
Therefore, effective multimodal learning algorithms are essential to integrate heterogeneous EHRs for better ICU outcome prediction. 

Recently, deep learning-based multimodal models have been proven to combine structured EHR data and free-text data at the deep feature level for clinical outcome predictions \cite {zhang2020combining, shin2021early, lin2024prediction}. 
These models often simply concatenate structured data with encoded features from free texts to generate patient representations for decision-making. While those approaches have improved prediction accuracy, the clinical impacts between the two heterogeneous modalities \cite{guo2019deep} are now well explored.
Another limitation of existing research is the lack of reliability evaluation for deep learning models. 
Unreliable predictions can lead to incorrect diagnoses or treatment plans, potentially harming patients \cite{kelly2019key,kumar2023artificial,wubineh2024exploring}. Therefore, evaluating the prediction reliability is crucial beyond just predictive accuracy, especially in critical care settings. However, concerns about the reliability of existing models in noisy and unstable clinical environments still remain.

Belief function theory (BFT), also known as Dempster-Shafer theory (DST), is a powerful framework for modeling, reasoning with, and integrating imperfect (noisy, uncertain, conflicting) data \cite{shafer1976mathematical, dempster1967upper, denoeux2000neural}. The effectiveness of BFT in low-quality and multimodal medical image analysis has been widely reviewed in \cite{huang2023application, huang2024review}. However, the study of BFT in EHR data is limited. Ling et al.\cite{huang2024evidential} first studied the survival prediction uncertainty using structured EHRs under the framework of BFT and possibility theory. The heterogeneity of clinical data and other medical modalities, e.g., imaging and genetic, are also studied in \cite{huang2024esurvfusion} by combining multimodal data using the generated Dempster's combination rule \cite{denoeux2023reasoning}. 

In this work \footnote{This work is an extended version of the short paper presented at the 8th International Conference on Belief Functions (BELIEF 2024) \cite{ruan2024evidence}.}, we further study the effectiveness of BFT in multimodal EHR analysis using structured EHRs and free-text notes with a focus on ICU outcomes prediction. 
We propose a multimodal learning model under the BFT framework with accurate and reliable ICU outcomes prediction using multimodal EHR data. Instead of developing more effective feature extraction or interaction strategies for multimodal data communication, our framework focuses on effective evidence fusion study and integrates information based on the evidence derived from different modalities. Specifically, we use state-of-the-art deep neural networks for single-modality feature extraction: ResNet/Transformer-based models for structured EHR data and pre-trained language models for free-text EHR data. The extracted features are independently mapped into evidence with an evidence mapping module and then combined in the evidence space in an evidence fusion module. 
Experimental results on the MIMIC-III database for mortality and prolonged length of stay (PLOS) predictions demonstrate the effectiveness of our proposed model in both predictive accuracy and reliability.

\section{Preliminaries}
\subsection{Belief function theory}
\label{subsec: dst}
Belief function theory (BFT) was first introduced by Dempster and Shafer \cite{dempster1967upper,shafer1976mathematical}. The expressive capabilities of belief functions enable a more accurate representation of evidence than relying solely on probabilities. Let $\Omega = \{\omega_1, \omega_2, \cdots, \omega_M\}$ be a finite set of hypotheses about some question, called the \textit{frame of discernment}. Evidence about a variable taking values in $\Omega$ can be represented by a \emph{mass function}: $2^{\Omega}$ to [0,1] such that $m(\emptyset) = 0$ and 
\begin{equation}
    \sum_{A \subseteq \Omega}m(A) = 1.
\end{equation} 
For any hypothesis $A \subseteq \Omega$, the quantity $m(A)$ is interpreted as a share of a unit mass of belief allocated to the hypothesis that the truth is in $A$, and which cannot be allocated to any strict subset of $A$ based on the available evidence. Set $A$ is called a \textit{focal set} of $m$ if $m(A) > 0$. A mass function is said to be Bayesian if its focal sets are singletons, and logical if it has only one focal set. Two mass functions $m_1$ and $m_2$ representing independent items of evidence can be combined conjunctively by \textit{Dempster’s combination rule} \cite{shafer1976mathematical} $\oplus$ as
\begin{equation}
    (m_{1} \oplus m_{2}) (A) = \frac{\sum_{B \cap C = A} m_1 (B) m_2 (C) }{1-\sum_{B \cap C = \emptyset} m_1 (B) m_2 (C)}, 
    \label{eq: dst}
\end{equation}
for all $A \neq \emptyset$, where $\sum_{B \cap C = \emptyset} m_1 (B) m_2 (C)$ is the degree of conflict among the two pieces of evidence, The nice information fusion attribute of BFT points out the high potential in heterogenetic medical data analysis. 

After aggregating all available evidence, the final decision of BFT can be made based on the pignistic transformation proposed by Smets in the Transferable Belief Model \cite{smets94a} that combinese all mass functions using the following expression:
\begin{equation}
    p(\omega) = \sum_{A \subseteq \Omega: \omega \in A} \frac{m(A)}{|A|}, \forall \omega \in \Omega.
    \label{eq: pignis}
\end{equation}

\subsection{Evidential neural network}
Den{\oe}ux \cite{denoeux2000neural} proposed an evidential neural network (ENN) that maps imperfect (uncertain, imprecise, or noise) input features into degrees of belief and ignorance (uncertainty) under the framework of BFT \cite{shafer1976mathematical, dempster2008upper}. 
The essential concept of ENN is to consider each prototype as a piece of evidence, which is discounted based on its distance from the input vector. The evidence from different prototypes is then aggregated using Dempster's combination rule. 

\begin{figure}[htb]
    \centering
    \includegraphics[width=0.9\linewidth]{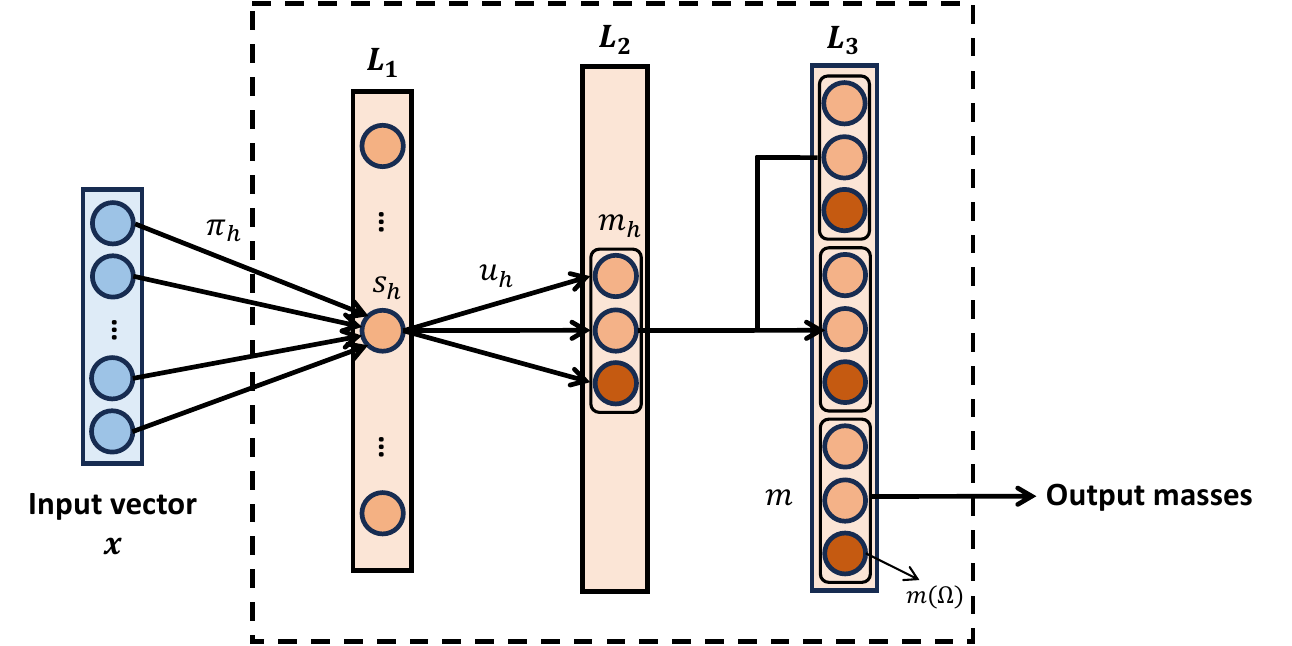}
    \caption{The illustration of ENN}
    \label{fig:enn}
\end{figure}

As illustrated in Figure \ref{fig:enn}, the ENN consists of one input layer, one hidden layer, and one output layer. The input layer is composed of $H$ units ($H$ is the number of prototypes), whose weights vectors are prototypes $\pi_{1}, \pi_{2}, \cdots, \pi_{H}$ in input space. 
The activation of unit $h$ in the input layer is
\begin{equation}
    s_{h} = \beta_{h} \mathrm{exp}(- \gamma_{h} d_{h}^2),
\end{equation}

where $d_{h} = ||x - \pi_{h} ||$ denotes the Euclidean distance between input vector $x$ and prototype $\pi_{h}$ , $\gamma_{h} > 0$ is a scale parameter, and $\beta_{h} \in [0, 1]$ is an extra parameter.
The hidden layer computes mass functions $m_{h}$ (evidence) of each prototype $\pi_{h}$ is defined as:
\begin{subequations}
\begin{align}
    m_{h} (\{ \omega_c\}) &= u_{h}^{(c)} s_{h}, \quad c = 1, 2, \cdots, M,\\
        m_{h} (\Omega) &= 1 - s_{h},
\end{align}
\end{subequations}
where $u_{h}^{(c)}$ is the membership degree of prototype $h$ to class $\omega_c$, $\sum_{c = 1}^M u_{h}^{(c)} = 1$, and $M$ is the number of classes. Therefore, the vector of mass functions induced by prototypes is denoted as:
\begin{equation}
    m_{h} = (m_{h} (\{ \omega_1\}), m_{h} (\{ \omega_2\}), \cdots, m_{h} (\{ \omega_M\}), m_{h} (\Omega))  \in \mathbb{R}^{M+1}.
\end{equation}
Finally, the mass functions are then aggregated by Dempster’s combination rule
using Eq. \ref{eq: dst} in the output layer. A combined mass function $m$ is computed as the orthogonal sum of the $H$ mass functions:

\begin{equation}
    m = m_{1} \oplus m_{2} \oplus \cdots \oplus m_{H} \in \mathbb{R}^{M+1}.
\end{equation}
The combined mass functions (the outputs of the ENN) represent the degrees of belief about the given class with $m(\{ \omega_c\})$, as well as its prediction uncertainty with $m(\Omega)$. In our binary classification case, the dimension of ENN outputs would be three.

\section{METHODS}

\subsection{Model architecture}
\begin{figure}[!htb]
    \centering
    \includegraphics[width=0.95\linewidth]{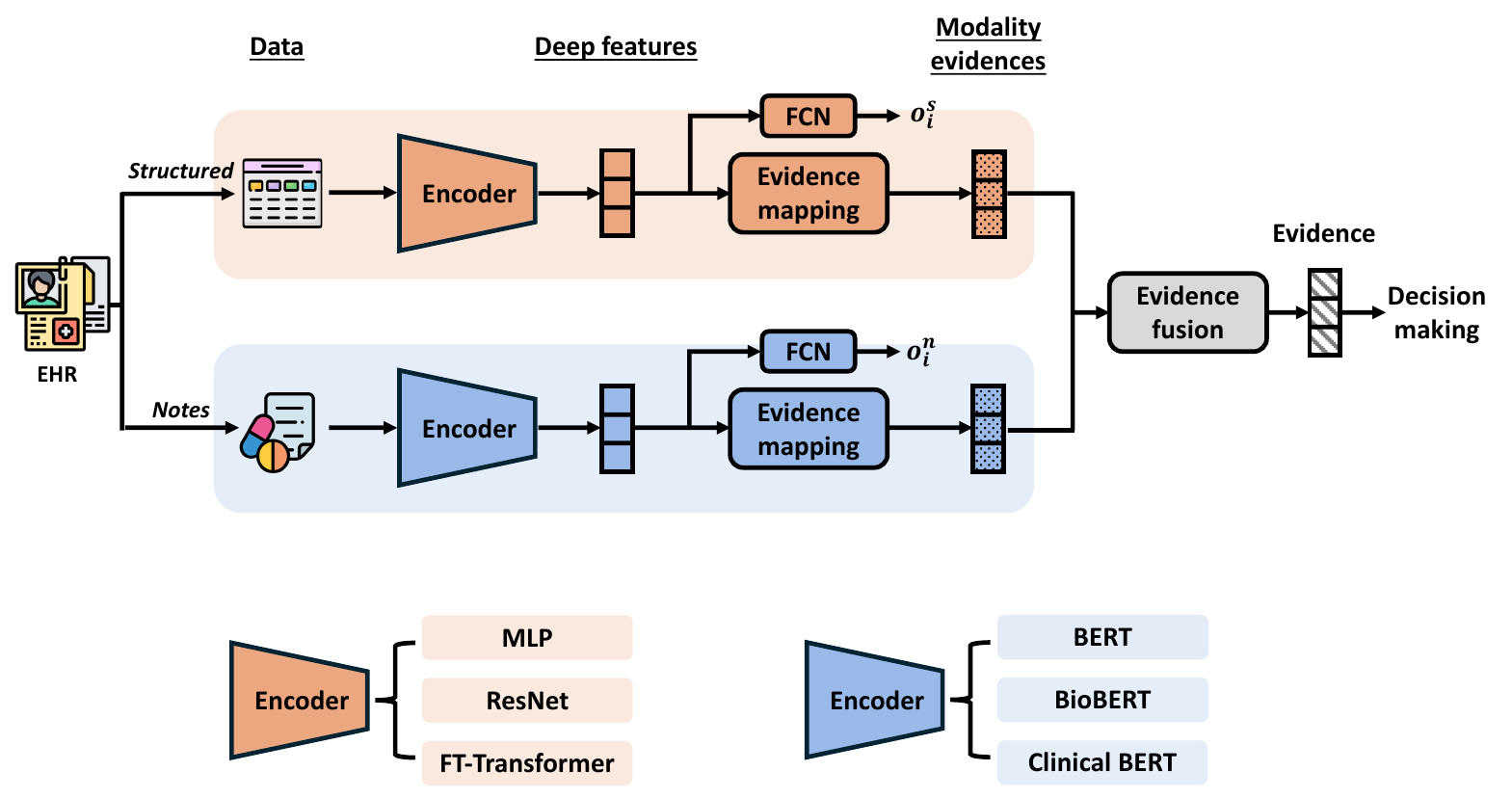}
    \caption{The overview of our proposed framework. EM: evidence mapping and EF: evidence fusion}
    \label{fig:overview}
\end{figure}

The overview of the proposed framework is illustrated in Figure \ref{fig:overview}. 
The general idea of this framework is to generate the modality-level evidence for both structured data and free-text notes using the evidence mapping module and fuse the modality-level evidence with Dempster's combination rule for final prediction. 

\subsubsection{Evidence mapping (EM)}

Inspired by ENN and its promising adoptions in medical data analysis \cite{lian2018joint, huang2022lymphoma, huang2022evidence, huang2025deep}, we propose incorporating ENN as an evidence mapping module with the state-of-the-art encoders to generate evidence for structured EHRs and free-text notes. 
Given modality level input, the evidence mapping module can output the evidence for each class as well as the uncertainty regarding this prediction.

\paragraph{Structured data evidence mapping}
To produce modality evidence for structured data, we initially used a structured data encoder to extract deep features (the output dimension is set to 32). We considered three popular encoders to extract the embeddings: MLP, ResNet, FT-Transformer \cite{gorishniy2021revisiting} (see 
\hyperref[encoder_details]{Baselines section} for more details).
Subsequently, we introduced an evidence mapping module (the number of prototypes is set to 20) to transform the deep features into evidence embeddings for structured data.

\paragraph{Free-text notes evidence mapping}
Similarly, we utilized pre-trained language models to extract the deep features from clinical notes, on which we developed an evidence mapping module (the number of prototypes is set to 20) to produce the evidence of the modality. Our primary analysis focused on pre-trained architectures similar to BERT. Accordingly, three BERT-based architectures were evaluated: BERT \cite{devlin2018bert}, BioBERT \cite{devlin2018bert}, Clinical BERT \cite{alsentzer2019publicly} (see \hyperref[encoder_details]{Baselines section} for more details). To minimize computational overhead while maintaining high predictive performance, we froze the pre-trained language models and fine-tuned an additional layer (128 hidden units) on top in model training.

\subsubsection{Evidence fusion (EF)}
Based on the modality-level evidence obtained from structured EHR and free-text notes, we developed the evidence fusion module based on Dempster's combination rule to generate the final evidence for decision-making. 

To combine multiple mass functions $m_1, m_2, \cdots, m_K$ from different modalities/data types/data sources, Dempster's combination rule is applied again Eq. \ref{eq: dst} to aggregate evidence for multiple sources for final evidence generation.

\begin{equation}
    m = m_{1} \oplus m_{2} \oplus \cdots \oplus m_{K} \in \mathbb{R}^{M+1},
\end{equation}
where $K$ is the number of mass functions to combine. For example, $K$=2 for the fusion of evidence from structured EHRs and free-text notes.

\subsection{Augmented model optimization algorithm}
We optimize the proposed framework using an augmented learning algorithm, which includes two types of optimization objectives: (1) main objective and (2) auxiliary objective. The main objective is to optimize predictive performance based on transformed evidence as the primary loss function. Additionally, two auxiliary cross-entropy losses are incorporated to enhance the feature representation capability of the independent encoders for the two modalities, as the evidence mapping module performs more effectively with high-quality representations.

Let $p_i = (p_i(\omega_1), \cdots, p_i (\omega_c), \cdots, p_i (\omega_M)) $ be the final probability after the pignistic transformation \eqref{eq: pignis} for training sample $i$, and $y_i = (y_{i, 1}, y_{i, 2}, \cdots, y_{i, M})$ denotes the one-hot encoding for corresponding ground-truth labels. 
The main loss function $\mathcal{L}_{main}$ is computed as:
\begin{equation}
    \mathcal{L}_{main} = - \frac{1}{N} \sum_{i=1}^{N} \sum_{c=1}^{M} w_c y_{i, c} \mathrm{log} (p_i (\omega_c)),
\end{equation}
where $N$ is the number of training samples, $M$ denotes the number of classes, and $w_c$ is the weight assigned to each class to address the class imbalance issue.

Moreover, two auxiliary cross-entropy losses are introduced to optimize the feature representation performance of the encoders, as the evidence mapping module performs more effectively with high-quality representations.
Firstly, to regulate the representation generated by encoders, we added an additional fully connected network (FCN) to generate logits $o_{i}$ for each modality. Let $o^s_{i} = (o_{i, 1}^{s}, o_{i, 2}^{s}, \cdots, o_{i, M}^{s})$ and $o^n_{i} = (o_{i, 1}^{n}, o_{i, 2}^{n}, \cdots, o_{i, M}^{n})$ be the logits from the encoders for structured data and free-text notes, respectively, the cross-entropy losses $\mathcal{L}^s_{aux}$ and $\mathcal{L}^n_{aux}$ are then calculated with $y_i$ for structured data and notes, respectively:

\begin{equation}
    \mathcal{L}^s_{aux}  = - \frac{1}{N} \sum_{i=1}^{N} \sum_{c=1}^{M} w_c y_{i, c} \mathrm{log} \frac{\mathrm{exp}(o_{i, c}^{s})}{\sum_{b=1}^{M} \mathrm{exp}(o_{i, b}^{s})},
\end{equation}

\begin{equation}
    \mathcal{L}^n_{aux}  = - \frac{1}{N} \sum_{i=1}^{N} \sum_{c=1}^{M} w_c y_{i, c} \mathrm{log} \frac{\mathrm{exp}(o_{i, c}^{n})}{\sum_{b=1}^{M} \mathrm{exp}(o_{i, b}^{n})},
\end{equation}

Ultimately, the overall loss function $\mathcal{L}_{overall}$ is defined as follows:
\begin{equation}
    \mathcal{L}_{overall} = \mathcal{L}_{main} + \alpha \mathcal{L}^s_{aux} + \beta \mathcal{L}^n_{aux},
\end{equation}
where $\alpha, \beta$ are the hyperparameters that control the balance between the main loss and the auxiliary cross-entropy losses. In both tasks, we set $\alpha = 2$ and $\beta = 1$.

\section{Experiments}


\subsection{Study Cohort}

This study used data from MIMIC-III (Medical Information Mart for Intensive Care III), a large, publicly available database containing de-identified health records from patients in critical care units at Beth Israel Deaconess Medical Center (from the US) between 2001 and 2012 \cite{johnson2016mimic}. We collected structured EHR data and free-text clinical notes from the database. Patients were excluded if they (1) were under 18 years of age at admission and (2) had incomplete length of stay or mortality data. For patients with multiple ICU stays, we considered only the first.

\subsection{Input features}
Input features include both structured EHR data and unstructured free-text notes. In this section, we demonstrate the patient features in our study and provide details about the data preprocessing steps.

\paragraph{Structured EHR data}
The structured data were collected during patients' ICU stays and included demographic information, vital signs and laboratory tests, medical treatments, and comorbidities. 
For demographic information, the patient's age, gender, weight, ethnicity, and admission type at the time of admission were included in the study. 
Vital signs/lab tests are the most crucial health indicators, easily measured using non-invasive equipment, and are readily understood by all healthcare professionals. For each variable considered, we used the first value recorded within 24 hours of admission time. We then excluded any variables with $\geq 50\%$  missingness rate; this resulted in the inclusion of only heart rate among vital signs features alongside 19 lab test features, such as blood urea nitrogen, eosinophil count, and lymphocyte count, over the same period. All vital sign/lab test features were numerical variables.
Medical treatments, which include services and interventions provided to patients and recorded in digital systems, were also analyzed. Treatments such as sedatives, statins, diuretics, antibiotics, ventilation, and vasopressors were included, with each treatment feature coded as a binary variable, indicating whether the patient received the treatment. 
Comorbidities refer to the presence of additional medical conditions, which play a role in decision-making models. In this study, comorbidities such as hypertension, diabetes, alcohol abuse, cerebrovascular accident (stroke), congestive heart failure, and ischemic heart disease were included, all represented as binary variables.

All categorical features were encoded using one-hot encoding, and numerical features were normalized. Missing data were addressed by imputing the mean for continuous features and the mode for categorical features, ensuring data consistency. Eventually, the structured data contained 41 features.

\paragraph{Free-text EHR notes}
Free-text notes contain a rich repository of clinical information about observations, assessments, and the overall clinical picture, which structured data often fails to capture. Furthermore, they provide an important context for interpreting structured data. For instance, while lab results may indicate abnormal values, free-text notes can clarify the relevance of these results by considering the patient's history, comorbidities, or specific circumstances at the time of testing. As a result, NLP techniques, especially pre-trained LLM, can be applied to these notes to gain deeper insights for data-driven predictions. In this study, we focus on \textit{Nursing}, \textit{Nursing/Other}, \textit{Physician}, and \textit{Radiology} notes, as these comprise the majority of clinical documentation and are frequently recorded in the MIMIC-III database \cite{zhang2020combining}. We extracted only the first 24 hours of notes for each admission to facilitate early outcome prediction.

All notes were preprocessed by appending the feature name at the front to help the pre-trained language model better understand the clinical texts. For instance, if the content \textit{[x]} of a note is under \textit{Nursing},  the processed note would be \textit{Nursing: [x]}.
The four types of notes were then concatenated using a newline symbol (\textbackslash n) to form a unified \textit{Notes} for each patient.
Tokenizers from pre-trained language models in Huggingface were employed to break the notes into tokens, standardizing the free-text data for further NLP tasks. The \textit{Notes} was transformed to a fixed length of 512 tokens to ensure input consistency; longer notes were truncated, while shorter notes were padded.

\subsection{Prediction Tasks}
In this study, we focus on two ICU prediction outcomes: mortality and prolonged length of stay. Since the two clinical outcomes are rare in patient popularity, we applied a simple class weighting approach during training based on relative class frequencies to mitigate biases and handle the imbalance in EHR data.

\paragraph{Mortality}
Mortality is widely acknowledged as a critical outcome in ICUs. The primary objective of this task is to determine whether a patient is likely to die during their hospital stay. Accurate predictions enable the early identification of high-risk patients and support the efficient allocation of ICU resources. This prediction task is typically framed as a binary classification problem, with the label indicating the occurrence or absence of a death event.

\paragraph{Prolonged length of stay}
Length of stay refers to the duration between a patient’s admission to and discharge from the ICU. In this study, we aim to predict prolonged length of stay (PLOS), defined as a stay exceeding 7 days \cite{liu2010length, rajkomar2018scalable, zhang2020combining}. Prolonged ICU stays are often linked to severe illnesses, complications, and increased mortality. Moreover, they place considerable pressure on hospital resources by reducing the availability of ICU beds and specialized personnel. Efficient management of ICU LOS not only improves patient outcomes but also enhances the overall effectiveness of healthcare systems. This problem is framed as a binary classification task.

\subsection{Baselines \label{encoder_details}}


To comprehensively evaluate the effectiveness of our proposed fusion framework, we compared it against three baseline model categories:
(1) models using only structured data, (2) models using only free-text notes, and (3) existing multimodal models that integrate both data types.

\paragraph{Structured data baseline }
The following models were used to evaluate performance with structured EHR data:

\begin{itemize}
\item \underline{Random Forest} \cite{breiman2001random}: A decision tree-based ensemble learning method for making predictions.

\item \underline{MLP} \cite{gorishniy2021revisiting}: A fundamental neural network with fully connected layers to encode structured data and serves as a reliable baseline. The model was configured with 3 layers, 32 hidden units, and a dropout rate of 0.1.

\item \underline{ResNet} \cite{gorishniy2021revisiting}: Because of the success of ResNet in computer vision \cite{he2016deep}, it has also been adapted for structured data modeling. Specifically, the main building block is simplified by providing a direct path from input to output. The configuration in our study has 3 residual blocks, 32 hidden units, and a dropout rate of 0.1.

\item \underline{FT-Transformer} \cite{gorishniy2021revisiting}: It converts all categorical and numerical features into embeddings, which are then processed through a couple of Transformer layers. It has demonstrated superior performance as a structured data encoder across various tasks. The model configuration has 3 Transformer layers with 192 hidden units, 8 attention heads, and a dropout rate of 0.2.
\end{itemize}

\paragraph{Free-text notes baseline }
To assess text-based prediction performance, we compared our model against three BERT-based text classification approaches:
\begin{itemize}


\item \underline{BERT} \cite{devlin2018bert}: A pre-trained language model trained on a large English corpus using self-supervised learning. It learns contextual representations through masked language modeling and next-sentence prediction.  In this study, we used the Google-bert/bert-base-uncased model from Huggingface \cite{wolf2019huggingface} for feature extraction.

\item  \underline{BioBERT} \cite{lee2020biobert}: a variant of BERT pre-trained on biomedical literature, such as PubMed abstracts, and is optimized to perform more effectively on biomedical NLP tasks. In this study, we used \textit{dmis-lab/biobert-v1.1} from Huggingface as the extraction model.

\item \underline{Clinical BERT} \cite{alsentzer2019publicly}:
A fine-tuned version of BERT on clinical notes from the MIMIC-III database, making it well-suited for handling medical terminology and clinical narratives to enhance performance on clinical tasks. In our study, we used \textit{emilyalsentzer/Bio\_ClinicalBERT} from the Huggingface transformer library for extracting embeddings from clinical notes.
\end{itemize}

\paragraph{Multimodal modal baseline}

To compare against multimodal models, we implemented the concatenation-based approach from \cite{lin2024prediction}. This method combines structured EHR data with extracted text embeddings from free-text notes, followed by two fully connected layers for prediction. To ensure fairness, we tested this approach with BERT, BioBERT, and Clinical BERT as text encoders.

\subsection{Implementation details}

The dataset was randomly divided with 60\% for training, 20\% for validation, and 20\% for testing. For model training, a mini-batch size of 32 was used, and the maximum number of epochs was set to 150, with early stopping applied. 
To handle data imbalance, we used a simple class weighting technique \footnote{\url{https://scikit-learn.org/stable/modules/generated/sklearn.utils.class_weight.compute_class_weight.html}} based on class frequencies, as this was not the main focus of our research. The positive to negative weight ratios were set to 4.254:0.567 for the mortality prediction task and 3.660:0.579 for the PLOS prediction task.

Each model was trained five times with different random seeds, and we reported the average results along with the standard error of the metrics to ensure statistical reliability. Hyperparameters were optimized for all baseline models to achieve the best results. All compared models were implemented using Scikit-learn \cite{pedregosa2011scikit}, PyTorch \cite{paszke2019pytorch}, and Hugging Face's Transformers library \cite{wolf2019huggingface} in Python 3.8.19. The MLP, ResNet, and FT-Transformer models were built using the original source code on GitHub \footnote{\url{https://github.com/yandex-research/rtdl-revisiting-models.}}. The Multimodal approach was modified based on the code on Github \footnote{\url{https://github.com/WeiChunLin/Bio_Clinical_BERT_Multimodal_Model.}}.

\subsection{Model evaluation}
For comprehensive model evaluation and comparison, we reported the two types of metrics: predictive accuracy and reliability.
\begin{itemize}
    \item \textbf{Predictive accuracy} ensures that models correctly identify critically ill patients who require urgent intervention, thereby reducing the risk of misdiagnosis and unnecessary treatments. To comprehensively assess accuracy, we consider two aspects:
    
\begin{itemize} 
\item \textbf{Class-specific} accuracy metrics which evaluate performance separately for positive and negative cases using metrics such as precision, recall, specificity, and negative predictive value (NPV). 

\begin{align}
\text{Precision} & = \frac{\text{TP}}{\text{TP} + \text{FP}}, \\
\text{Recall} & = \frac{\text{TP}}{\text{TP} + \text{FN}}, \\
\text{Specificity} & = \frac{\text{TN}}{\text{TN} + \text{FP}}, \\
\text{NPV} & = \frac{\text{TN}}{\text{TN} + \text{FN}},
\end{align}
where TP, TN, FP, and FN are True Positive, True Negative, False Positive, and False Negative, respectively.

\item \textbf{Holistic} accuracy metrics include balanced accuracy (BACC), F1 score, the area under the receiver operating characteristic curve (AUROC), and the area under the precision-recall curve (AUPRC). AUROC is determined by calculating the area under the ROC curve (TP rate against FP rate across different threshold settings) and AUPRC calculates the area under the Precision-Recall curve across various threshold settings.
\begin{align}
    \text{BACC} &= \frac{1}{2} \left( \text{Recall} + \text{Specificity} \right), \\
    \text{F1} &= 2 \cdot \frac{\text{Precision} \cdot \text{Recall}}{\text{Precision} + \text{Recall}}.
\end{align}
BACC, F1 score, and AUPRC are well-suited for evaluating model performance in imbalanced class distributions, as they provide a more balanced assessment than traditional accuracy measures.
\end{itemize}
    
\item \textbf{Reliability} metrics quantify the model's confidence in predictions. A model with high reliability provides well-calibrated probability estimates, allowing clinicians to make informed risk assessments. Here, we evaluate the performance of the Brier score and negative log-likelihood (NLL). 

\begin{align}
\text{Brier Score} &= \frac{1}{N} \sum_{i=1}^{N} (p_i - o_i)^2, \\
\text{NLL} &= -\frac{1}{N} \sum_{i=1}^{N} \left( o_i \log(p_i) + (1 - o_i) \log(1 - p_i) \right),
\end{align}
where $N$ is the number of instances, $p_i$ is the predicted probability of the positive class for instance $i$, and $o_i$ is the actual outcome for instance $i$ (1 if positive, 0 if negative).
\end{itemize}

\section{RESULTS}

\subsection{Data description}
Table \ref{tab: characteristics} shows the descriptive characteristics of the patients in the study cohort. Our cohort includes 38469 patients in total under the inclusion criteria, in which 4540 (11.8\%) patients were identified as dead during the stay while 5220 (13.6\%) patients were identified as having prolonged length of stay. The patient demographic showed that the majority of patients were white, and male patients were slightly more than the female. Most of the admitted patients in the ICU were identified as emergency. Over half (51.2\%) of patients received mechanical ventilation during their ICU stay and hypertension (38.8\%) and congestive heart failure (31.2\%) were among the most common patient comorbidities. 

\begin{table}[htbp]
\centering
\begin{tabular}{p{2.3cm}p{2.1cm}p{2.2cm}p{2.1cm}p{2.2cm}}
\toprule
\multirow{2}{*}{\textbf{\shortstack[l]{Patient \\ characteristic}}} & \multicolumn{2}{l}{\textbf{Mortality}}        & \multicolumn{2}{l}{\textbf{PLOS}}             \\ 
\cmidrule{2-5} 
                                                & \textbf{Yes (N=4540)} & \textbf{No (N=33928)} & \textbf{Yes (N=5220)} & \textbf{No (N=33248)} \\ \midrule
Age                                             & 68.6 (15.1)           & 61.6 (16.9)            & 62.9 (16.3)           & 62.3 (16.9)            \\
Gender                                          &                       &                       &                       &                       \\
\qquad Male                                            & 2399 (52.8\%)         & 19378 (57.1\%)        & 2943 (56.4\%)         & 18834 (56.6\%)        \\
\qquad Female                                          & 2141 (47.2\%)         & 14550 (42.9\%)        & 2277 (43.6\%)         & 14414 (43.4\%)        \\
Weight                                          & 79.1 (22.7)           & 83.0 (23.1)           & 84.6  (24.3)          & 82.2 (22.9)           \\
Ethnicity                                       &                       &                       &                       &                       \\
\qquad White                                           & 3100 (68.3\%)         & 24343 (71.7\%)        & 3661 (70.1\%)         & 23782 (71.5\%)        \\
\qquad Black                                           & 260 (5.7\%)           & 2688 (7.9\%)          & 354 (6.8\%)           & 2594 (7.8\%)          \\
\qquad Asian                                           & 106 (2.3\%)           & 1166 (3.4\%)          & 153 (2.9\%)           & 1101 (3.3\%)          \\
\qquad Hispanic                                        & 88 (1.9\%)            & 803 (2.4\%)           & 104 (2.0\%)           & 805 (2.4\%)           \\
\qquad Other                                  & 986 (21.7\%)          & 4928 (14.5\%)         & 948 (18.2\%)          & 4966 (14.9\%)         \\
Admission type                                  &                       &                       &                       &                       \\
\qquad Emergency                                       & 4240 (93.4\%)         & 27062 (79.8\%)        & 4480 (85.8\%)         & 26822 (80.7\%)        \\
\qquad Elective                                        & 165 (3.6\%)           & 5911 (17.4\%)         & 533 (10.2\%)          & 5543 (16.7\%)         \\
\qquad Urgent                                         & 135 (3.0\%)           & 955 (2.8\%)           & 207 (4.0\%)           & 883 (2.7\%)           \\
Heart rate                                      & 91.4 (22.0)           & 86.8 (18.7)           & 91.5 (21.0)           & 86.7 (18.8)           \\
APTT                                            & 39.7 (27.1)           & 34.7 (21.6)           & 37.2 (23.8)           & 35.0 (22.2)           \\
BUN                                             & 34.1 (25.1)           & 23.8 (19.0)           & 28.1 (22.3)           & 24.5 (19.7)           \\
Eosinophil                                      & 1.3 (3.1)             & 1.5 (1.9)             & 1.2 (1.7)             & 1.5 (2.1)             \\
Lymphocytes                                     & 12.1 (11.9)           & 15.5 (11.4)           & 12.5 (10.8)           & 15.4 (11.6)           \\
Neutrophils                                     & 77.3 (17.7)           & 76.6 (13.9)           & 77.8 (15.2)           & 76.5 (14.4)           \\
RDW                                             & 15.5 (2.4)            & 14.5 (1.9)            & 14.9 (2.1)            & 14.6 (2.0)            \\
Bicarbonate                                     & 22.6 (5.7)            & 24.3 (4.4)            & 23.6 (5.2)            & 24.1 (4.5)            \\
Chloride                                        & 103.3 (7.3)           & 104.1 (6.0)           & 104.0 (6.7)           & 104.0 (6.1)           \\
Creatinine                                      & 1.6 (1.4)             & 1.3 (1.4)             & 1.4 (1.5)             & 1.3 (1.4)             \\
Hemoglobin                                      & 11.3 (2.3)            & 11.8 (2.3)            & 11.6 (2.2)            & 11.7 (2.3)            \\
Mean cell volume                                & 91.5 (7.7)            & 89.4 (6.6)            & 90.3 (7.1)            & 89.6 (6.7)            \\
Platelet count                                  & 227.2 (132.2)         & 239.0 (112.3)         & 231.8 (119.8)         & 238.5 (114.0)         \\
Potassium                                       & 4.3 (0.9)             & 4.2 (0.7)             & 4.2 (0.8)             & 4.2 (0.7)             \\
Sodium                                          & 138.2 (6.2)           & 138.6 (4.6)           & 138.7 (5.2)           & 138.5 (4.7)           \\
PT                                              & 17.4 (10.6)           & 15.1 (6.8)            & 16.0 (8.2)            & 15.3 (7.3)            \\
INR                                             & 1.8 (2.2)             & 1.4 (1.3)             & 1.6 (1.9)             & 1.4 (1.4)             \\
WBC                                             & 14.3 (16.5)           & 11.5 (8.8)            & 12.9 (8.8)            & 11.6 (10.2)           \\
PLR                                             & 39.9 (56.7)           & 30.4 (47.4)           & 38.4 (57.5)           & 30.6 (47.3)           \\
NLR                                             & 13.9 (15.6)           & 9.9 (13.7)            & 13.0 (15.2)           & 10.1 (13.8)           \\
Sedatives                                       & 1263 (27.8\%)         & 9114 (26.9\%)         & 2360 (45.2\%)         & 8017 (24.1\%)         \\
Statin                                          & 405 (8.9\%)           & 5164 (15.2\%)         & 556 (10.7\%)          & 5013 (15.1\%)         \\
Diuretic                                        & 567 (12.5\%)          & 5435 (16.0\%)         & 865 (16.6\%)          & 5137 (15.5\%)         \\
Antibiotics                                     & 952 (21.0\%)          & 4836 (14.3\%)         & 1122 (21.5\%)         & 4666 (14.0\%)         \\
Ventilation                                     & 2326 (51.2\%)         & 10923 (32.2\%)        & 3105 (59.5\%)         & 10144 (30.5\%)        \\
Vasopressor                                     & 1538 (33.9\%)         & 6340 (18.7\%)         & 1639 (31.4\%)         & 6239 (18.8\%)         \\
Hypertension                                    & 1760 (38.8\%)         & 16221 (47.8\%)        & 2199 (42.1\%)         & 15782 (47.5\%)        \\
Diabetes                                        & 1107 (24.4\%)         & 9249 (27.3\%)         & 1401 (26.8\%)         & 8955 (26.9\%)         \\
Alcohol abuse                                   & 202 (4.4\%)           & 1554 (4.6\%)          & 333 (6.4\%)           & 1423 (4.3\%)          \\
CVA                                             & 309 (6.8\%)           & 1139 (3.4\%)          & 364 (7.0\%)           & 1084 (3.3\%)          \\
CHF                                             & 1416 (31.2\%)         & 8711 (25.7\%)         & 1867 (35.8\%)         & 8260 (24.8\%)         \\
IHD                                             & 1276 (28.1\%)         & 12390 (36.5\%)        & 1634 (31.3\%)         & 12032 (36.2\%)        \\ \bottomrule
\end{tabular}
\caption{Characteristics of structured features in the patient cohort. For categorical features, the number of instances in each category is reported along with the percentage. For continuous features, the mean and standard deviation are reported in the study.}
\label{tab: characteristics}
\end{table}

\subsection{Model performance}
\subsubsection{Predictive accuracy}
Table \ref{tab:model_performance_pred_mortality} and \ref{tab:model_performance_pred_plos} present the comparison of model performance for holistic predictive accuracy in mortality and PLOS prediction tasks, respectively.

\begin{table}[htp]
    \centering
    \begin{tabular}{p{2.3cm}p{0.8cm}p{0.6cm}p{1.6cm}p{1.6cm}p{1.65cm}p{1.6cm}}
\toprule
\textbf{Model}    & \textbf{Struct.} & \textbf{Notes} & \textbf{BACC$\uparrow$}                                        & \textbf{F1$\uparrow$}                                          & \textbf{AUROC$\uparrow$}                                       & \textbf{AUPRC$\uparrow$}                                       \\ \hline
Random Forest     & x                &                & 0.7172{\tiny{$\pm$0.0019}}                  & 0.3820{\tiny{$\pm$0.0018}}                  & 0.7988{\tiny{$\pm$0.0010}}                  & 0.3766{\tiny{$\pm$0.0030}}                  \\
MLP               & x                &                & 0.7486{\tiny{$\pm$0.0003}}                  & 0.4043{\tiny{$\pm$0.0019}}                  & 0.8326{\tiny{$\pm$0.0011}}                  & 0.4429{\tiny{$\pm$0.0033}}                  \\
ResNet            & x                &                & 0.7521{\tiny{$\pm$0.0011}}                  & 0.4113{\tiny{$\pm$0.0007}}                  & 0.8350{\tiny{$\pm$0.0006}}                  & 0.4468{\tiny{$\pm$0.0016}}                  \\
FT-Transformer    & x                &                & 0.7592{\tiny{$\pm$0.0025}}                  & 0.4166{\tiny{$\pm$0.0021}}                  & 0.8426{\tiny{$\pm$0.0013}}                  & 0.4577{\tiny{$\pm$0.0029}}                  \\ \hline
\multicolumn{7}{c}{BERT as text encoder}                                                                                                                                                                                                                                                                                  \\ \hline
Text encoder only &                  & x              & 0.6300{\tiny{$\pm$0.0014}}                  & 0.2814{\tiny{$\pm$0.0019}}                  & 0.6777{\tiny{$\pm$0.0014}}                  & 0.2037{\tiny{$\pm$0.0012}}                  \\
Multimodal        & x                & x              & 0.7531{\tiny{$\pm$0.0016}}                  & 0.4079{\tiny{$\pm$0.0031}}                  & 0.8398{\tiny{$\pm$0.0006}}                  & 0.4596{\tiny{$\pm$0.0021}}                  \\
\hdashline[2.5pt/5pt]
Ours (MLP)         & x                & x              & 0.7610{\tiny{$\pm$0.0015}}                  & \textbf{0.4507{\tiny{$\pm$0.0007}}}                  & \textbf{0.8486\tiny$\pm$0.0011} & 0.4796{\tiny{$\pm$0.0027}}                  \\
Ours (ResNet)     & x                & x              & 0.7581{\tiny{$\pm$0.0037}}                  & 0.4404{\tiny{$\pm$0.0038}}                  & 0.8415{\tiny{$\pm$0.0011}}                  & 0.4688{\tiny{$\pm$0.0014}}                  \\
Ours (FT-Trans)   & x                & x              & \textbf{0.7634\tiny$\pm$0.0012} & 0.4432{\tiny{$\pm$0.0060}}                  & 0.8485{\tiny{$\pm$0.0007}}                  & \textbf{0.4797{\tiny{$\pm$0.0029}}}                  \\ \hline
\multicolumn{7}{c}{BioBERT as text encoder}                                                                                                                                                                                                                                                                               \\ \hline
Text encoder only &                  & x              & 0.6164{\tiny{$\pm$0.0043}}                  & 0.2759{\tiny{$\pm$0.0021}}                  & 0.6695{\tiny{$\pm$0.0016}}                  & 0.2181{\tiny{$\pm$0.0009}}                  \\
Multimodal        & x                & x              & 0.7558{\tiny{$\pm$0.0007}}                  & 0.4073{\tiny{$\pm$0.0027}}                  & 0.8362{\tiny{$\pm$0.0006}}                  & 0.4570{\tiny{$\pm$0.0032}}                  \\
\hdashline[2.5pt/5pt]
Ours (MLP)         & x                & x              & 0.7492{\tiny{$\pm$0.0042}}                  & 0.4493{\tiny{$\pm$0.0060}}                  & 0.8408{\tiny{$\pm$0.0012}}                  & 0.4758{\tiny{$\pm$0.0009}}                  \\
Ours (ResNet)     & x                & x              & 0.7546{\tiny{$\pm$0.0039}}                  & 0.4438{\tiny{$\pm$0.0053}}                  & 0.8424{\tiny{$\pm$0.0012}}                  & 0.4671{\tiny{$\pm$0.0014}}                  \\
Ours (FT-Trans)   & x                & x              & \textbf{0.7610{\tiny{$\pm$0.0015}}}                  & \textbf{0.4507{\tiny{$\pm$0.0007}}}                  & \textbf{0.8486\tiny$\pm$0.0011} & \textbf{0.4796{\tiny{$\pm$0.0027}}}                  \\ \hline
\multicolumn{7}{c}{Clinical BERT as text encoder}                                                                                                                                                                                                                                                                         \\ \hline
Text encoder only &                  & x              & 0.6614{\tiny{$\pm$0.0007}}                  & 0.3080{\tiny{$\pm$0.0009}}                  & 0.7240{\tiny{$\pm$0.0001}}                  & 0.2928{\tiny{$\pm$0.0008}}                  \\
Multimodal        & x                & x              & 0.7584{\tiny{$\pm$0.0005}}                  & 0.4218{\tiny{$\pm$0.0019}}                  & 0.8404{\tiny{$\pm$0.0004}}                  & 0.4686{\tiny{$\pm$0.0026}}                  \\
\hdashline[2.5pt/5pt]
Ours (MLP)         & x                & x              & 0.7467{\tiny{$\pm$0.0022}}                  & \cellcolor{mycolor} 
\textbf{0.4629{\tiny{$\pm$0.0033}}}         & 0.8465{\tiny{$\pm$0.0008}}                  & 0.4935\tiny$\pm$0.0011 \\
Ours (ResNet)     & x                & x              & 0.7580{\tiny{$\pm$0.0023}}                  & 0.4472{\tiny{$\pm$0.0042}}                  & 0.8474{\tiny{$\pm$0.0006}}                  & 0.4899{\tiny{$\pm$0.0015}}                  \\
Ours (FT-Trans)   & x                & x              & \cellcolor{mycolor} \textbf{0.7672{\tiny{$\pm$0.0032}}}         & 0.4541\tiny$\pm$0.0032 & \cellcolor{mycolor} \textbf{0.8534{\tiny{$\pm$0.0012}}}         & \cellcolor{mycolor} \textbf{0.4977{\tiny{$\pm$0.0011}}}         \\ \hline
\end{tabular}
    \caption{Comparison of \textbf{predictive accuracy} on \textbf{mortality} prediction. The best results among models using the same text encoder are in bold, and the overall best results are shaded in grey.}
    \label{tab:model_performance_pred_mortality}
\end{table}

\begin{table}[t]
\begin{tabular}{p{2.3cm}p{0.8cm}p{0.6cm}p{1.6cm}p{1.6cm}p{1.65cm}p{1.6cm}}
\hline
\textbf{Model}    & \textbf{Struct.} & \textbf{Notes} & \textbf{BACC$\uparrow$}                                        & \textbf{F1$\uparrow$}                                          & \textbf{AUROC$\uparrow$}                                       & \textbf{AUPRC$\uparrow$}                                       \\ \hline
Random Forest     & x                &                & 0.6680{\tiny{$\pm$0.0025}}                  & 0.3492{\tiny{$\pm$0.0038}}                  & 0.7270{\tiny{$\pm$0.0018}}                  & 0.2980{\tiny{$\pm$0.0029}}                  \\
MLP               & x                &                & 0.6845{\tiny{$\pm$0.0009}}                  & 0.3736{\tiny{$\pm$0.0011}}                  & 0.7528{\tiny{$\pm$0.0010}}                  & 0.3353{\tiny{$\pm$0.0010}}                  \\
ResNet            & x                &                & 0.6851{\tiny{$\pm$0.0017}}                  & 0.3774{\tiny{$\pm$0.0018}}                  & 0.7532{\tiny{$\pm$0.0004}}                  & 0.3366{\tiny{$\pm$0.0021}}                  \\
FT-Transformer    & x                &                & 0.6956{\tiny{$\pm$0.0011}}                  & 0.3790{\tiny{$\pm$0.0021}}                  & 0.7674{\tiny{$\pm$0.0006}}                  & 0.3404{\tiny{$\pm$0.0021}}                  \\ \hline
\multicolumn{7}{c}{BERT as text encoder}                                                                                                                                                                                                                                                                                  \\ \hline
Text encoder only &                  & x              & 0.6087{\tiny{$\pm$0.0011}}                  & 0.2960{\tiny{$\pm$0.0016}}                  & 0.6507{\tiny{$\pm$0.0015}}                  & 0.2262{\tiny{$\pm$0.0009}}                  \\
Multimodal        & x                & x              & 0.6903{\tiny{$\pm$0.0018}}                  & 0.3739{\tiny{$\pm$0.0012}}                  & 0.7625{\tiny{$\pm$0.0006}}                  & 0.3427{\tiny{$\pm$0.0007}}                  \\
\hdashline[2.5pt/5pt]
Ours (MLP)        & x                & x              & 0.6878{\tiny{$\pm$0.0016}}                  & 0.3868{\tiny{$\pm$0.0030}}                  & 0.7580{\tiny{$\pm$0.0004}}                  & 0.3499{\tiny{$\pm$0.0015}}                  \\
Ours (ResNet)     & x                & x              & 0.6931{\tiny{$\pm$0.0013}}                  & \textbf{0.3887{\tiny{$\pm$0.0040}}}                  & 0.7655{\tiny{$\pm$0.0011}}                  & \textbf{0.3584\tiny$\pm$0.0007} \\
Ours (FT-Trans)   & x                & x              & \textbf{0.7005\tiny$\pm$0.0010} & 0.3809{\tiny{$\pm$0.0024}}                  & \textbf{0.7725\tiny$\pm$0.0006} & 0.3524{\tiny{$\pm$0.0011}}                  \\ \hline
\multicolumn{7}{c}{BioBERT as text encoder}                                                                                                                                                                                                                                                                               \\ \hline
Text encoder only &                  & x              & 0.6057{\tiny{$\pm$0.0016}}                  & 0.2940{\tiny{$\pm$0.0012}}                  & 0.6487{\tiny{$\pm$0.0014}}                  & 0.2214{\tiny{$\pm$0.0012}}                  \\
Multimodal        & x                & x              & 0.6905{\tiny{$\pm$0.0022}}                  & 0.3731{\tiny{$\pm$0.0025}}                  & 0.7606{\tiny{$\pm$0.0007}}                  & 0.3364{\tiny{$\pm$0.0009}}                  \\
\hdashline[2.5pt/5pt]
Ours (MLP)        & x                & x              & 0.6866{\tiny{$\pm$0.0022}}                  & 0.3905{\tiny{$\pm$0.0038}}                  & 0.7573{\tiny{$\pm$0.0010}}                  & 0.3461{\tiny{$\pm$0.0023}}                  \\
Ours (ResNet)     & x                & x              & 0.6896{\tiny{$\pm$0.0011}}                  & \textbf{0.3918{\tiny{$\pm$0.0037}}}                  & 0.7647{\tiny{$\pm$0.0009}}                  & \textbf{0.3532{\tiny{$\pm$0.0010}}}                  \\
Ours (FT-Trans)   & x                & x              & \textbf{0.6986{\tiny{$\pm$0.0011}}}                  & 0.3847{\tiny{$\pm$0.0030}}                  & \textbf{0.7701{\tiny{$\pm$0.0012}}}                  & 0.3504{\tiny{$\pm$0.0014}}                  \\ \hline
\multicolumn{7}{c}{Clinical BERT as text encoder}                                                                                                                                                                                                                                                                         \\ \hline
Text encoder only &                  & x              & 0.6420{\tiny{$\pm$0.0016}}                  & 0.3261{\tiny{$\pm$0.0015}}                  & 0.6946{\tiny{$\pm$0.0009}}                  & 0.2661{\tiny{$\pm$0.0006}}                  \\
Multimodal        & x                & x              & 0.6933{\tiny{$\pm$0.0014}}                  & 0.3714{\tiny{$\pm$0.0023}}                  & 0.7648{\tiny{$\pm$0.0009}}                  & 0.3544{\tiny{$\pm$0.0018}}                  \\
\hdashline[2.5pt/5pt]
Ours (MLP)        & x                & x              & 0.6870{\tiny{$\pm$0.0012}}                  & 0.3952{\tiny{$\pm$0.0032}}                  & 0.7633{\tiny{$\pm$0.0008}}                  & 0.3557{\tiny{$\pm$0.0021}}                  \\
Ours (ResNet)     & x                & x              & 0.6974{\tiny{$\pm$0.0011}}                  & \cellcolor{mycolor} \textbf{0.4019{\tiny{$\pm$0.0015}}}         & 0.7705{\tiny{$\pm$0.0009}}                  & \cellcolor{mycolor} \textbf{0.3639{\tiny{$\pm$0.0010}}}         \\
Ours (FT-Trans)   & x                & x              & \cellcolor{mycolor} \textbf{0.7027{\tiny{$\pm$0.0011}}}         & 0.3942\tiny$\pm$0.0020 & \cellcolor{mycolor} \textbf{0.7743{\tiny{$\pm$0.0006}}}         & 0.3575{\tiny{$\pm$0.0014}}                  \\ \hline
\end{tabular}
\caption{Comparison of \textbf{predictive accuracy} on \textbf{PLOS} prediction. The best results among models using the same text encoder are in bold, and the overall best results are shaded in grey.}
    \label{tab:model_performance_pred_plos}
\end{table}

\begin{table}[htpb]


\begin{tabular}{p{2.3cm}p{0.8cm}p{0.6cm}p{1.65cm}p{1.6cm}}
\hline
\textbf{Model}    & \textbf{Struct.} & \textbf{Notes} & \textbf{Brier$\downarrow$}                                     & \textbf{NLL$\downarrow$}                                       \\ \hline
Random Forest     & x                &                & 0.1891{\tiny{$\pm$0.0006}}                  & 0.5639{\tiny{$\pm$0.0015}}                  \\
MLP               & x                &                & 0.1677{\tiny{$\pm$0.0015}}                  & 0.4937{\tiny{$\pm$0.0037}}                  \\
ResNet            & x                &                & 0.1654{\tiny{$\pm$0.0006}}                  & 0.4895{\tiny{$\pm$0.0015}}                  \\
FT-Transformer    & x                &                & 0.1648{\tiny{$\pm$0.0014}}                  & 0.4832{\tiny{$\pm$0.0037}}                  \\ \hline
\multicolumn{5}{c}{BERT as text encoder}                                                                                                                                                \\ \hline
Text encoder only &                  & x              & 0.2302{\tiny{$\pm$0.0070}}                  & 0.6523{\tiny{$\pm$0.0159}}                  \\
Multimodal        & x                & x              & 0.1691{\tiny{$\pm$0.0030}}                  & 0.4992{\tiny{$\pm$0.0077}}                  \\
\hdashline[2.5pt/5pt]
Ours (MLP)        & x                & x              & \textbf{0.1345{\tiny{$\pm$0.0050}}}                  & \textbf{0.4041{\tiny{$\pm$0.0128}}}                  \\
Ours (ResNet)     & x                & x              & 0.1435{\tiny{$\pm$0.0052}}                  & 0.4296{\tiny{$\pm$0.0133}}                  \\
Ours (FT-Trans)   & x                & x              & 0.1450{\tiny{$\pm$0.0049}}                  & 0.4345{\tiny{$\pm$0.0129}}                  \\ \hline
\multicolumn{5}{c}{BioBERT as text encoder}                                                                                                                                             \\ \hline
Text encoder only &                  & x              & 0.2214{\tiny{$\pm$0.0021}}                  & 0.6326{\tiny{$\pm$0.0048}}                  \\
Multimodal        & x                & x              & 0.1725{\tiny{$\pm$0.0028}}                  & 0.5084{\tiny{$\pm$0.0088}}                  \\
\hdashline[2.5pt/5pt]
Ours (MLP)        & x                & x              & \textbf{0.1301\tiny$\pm$0.0062} & \textbf{0.3927\tiny$\pm$0.0159} \\
Ours (ResNet)     & x                & x              & 0.1392{\tiny{$\pm$0.0061}}                  & 0.4187{\tiny{$\pm$0.0157}}                  \\
Ours (FT-Trans)   & x                & x              & 0.1358{\tiny{$\pm$0.0014}}                  & 0.4095{\tiny{$\pm$0.0045}}                  \\ \hline
\multicolumn{5}{c}{Clinical BERT as text encoder}                                                                                                                                       \\ \hline
Text encoder only &                  & x              & 0.2132{\tiny{$\pm$0.0030}}                  & 0.6086{\tiny{$\pm$0.0070}}                  \\
Multimodal        & x                & x              & 0.1606{\tiny{$\pm$0.0022}}                  & 0.4792{\tiny{$\pm$0.0046}}                  \\
\hdashline[2.5pt/5pt]
Ours (MLP)        & x                & x              & \cellcolor{mycolor} \textbf{0.1176{\tiny{$\pm$0.0027}}}         & \cellcolor{mycolor} \textbf{0.3594{\tiny{$\pm$0.0066}}}         \\
Ours (ResNet)     & x                & x              & 0.1387{\tiny{$\pm$0.0043}}                  & 0.4178{\tiny{$\pm$0.0110}}                  \\
Ours (FT-Trans)   & x                & x              & 0.1390{\tiny{$\pm$0.0043}}                  & 0.4192{\tiny{$\pm$0.0112}}                  \\ \hline
\end{tabular}

\caption{Comparison of \textbf{reliability} performance on \textbf{mortality} prediction.}
\label{tab:model_performance_rel_mortality}
\end{table}

\begin{table}[!h]


\begin{tabular}{p{2.3cm}p{0.8cm}p{0.6cm}p{1.65cm}p{1.6cm}}
\hline
\textbf{Model}    & \textbf{Struct.} & \textbf{Notes} & \textbf{Brier$\downarrow$}                             & \textbf{NLL$\downarrow$}                               \\ \hline
Random Forest     & x                &                & 0.2350{\tiny{$\pm$0.0009}}          & 0.6649{\tiny{$\pm$0.0018}}          \\
MLP               & x                &                & 0.1950{\tiny{$\pm$0.0018}}          & 0.5739{\tiny{$\pm$0.0044}}          \\
ResNet            & x                &                & 0.1929{\tiny{$\pm$0.0012}}          & 0.5702{\tiny{$\pm$0.0027}}          \\
FT-Transformer    & x                &                & 0.1947{\tiny{$\pm$0.0031}}          & 0.5704{\tiny{$\pm$0.0079}}          \\ \hline
\multicolumn{5}{c}{BERT as text encoder}                                                                                                                                \\ \hline
Text encoder only &                  & x              & 0.2316{\tiny{$\pm$0.0043}}          & 0.6557{\tiny{$\pm$0.0091}}          \\
Multimodal        & x                & x              & 0.2017{\tiny{$\pm$0.0028}}          & 0.5960{\tiny{$\pm$0.0082}}          \\
\hdashline[2.5pt/5pt]
Ours (MLP)        & x                & x              &  \textbf{0.1773{\tiny{$\pm$0.0022}}}          & \textbf{0.5288{\tiny{$\pm$0.0063}}}          \\
Ours (ResNet)     & x                & x              & 0.1815{\tiny{$\pm$0.0045}}          & 0.5398{\tiny{$\pm$0.0107}}          \\
Ours (FT-Trans)   & x                & x              & 0.1927{\tiny{$\pm$0.0039}}          & 0.5617{\tiny{$\pm$0.0100}}          \\ \hline
\multicolumn{5}{c}{BioBERT as text encoder}                                                                                                                             \\ \hline
Text encoder only &                  & x              & 0.2297{\tiny{$\pm$0.0029}}          & 0.6518{\tiny{$\pm$0.0063}}          \\
Multimodal        & x                & x              & 0.1974{\tiny{$\pm$0.0026}}          & 0.5888{\tiny{$\pm$0.0043}}          \\
\hdashline[2.5pt/5pt]
Ours (MLP)        & x                & x              & \textbf{0.1720{\tiny{$\pm$0.0031}}}          & \textbf{0.5172{\tiny{$\pm$0.0076}}}          \\
Ours (ResNet)     & x                & x              & 0.1752{\tiny{$\pm$0.0048}}          & 0.5241{\tiny{$\pm$0.0123}}          \\
Ours (FT-Trans)   & x                & x              & 0.1832{\tiny{$\pm$0.0045}}          & 0.5376{\tiny{$\pm$0.0113}}          \\ \hline
\multicolumn{5}{c}{Clinical BERT as text encoder}                                                                                                                       \\ \hline
Text encoder only &                  & x              & 0.2225{\tiny{$\pm$0.0031}}          & 0.6350{\tiny{$\pm$0.0069}}          \\
Multimodal        & x                & x              & 0.2027{\tiny{$\pm$0.0056}}          & 0.5871{\tiny{$\pm$0.0136}}          \\
\hdashline[2.5pt/5pt]
Ours (MLP)        & x                & x              & \cellcolor{mycolor} \textbf{0.1637{\tiny{$\pm$0.0054}}} & \cellcolor{mycolor} \textbf{0.4943{\tiny{$\pm$0.0145}}} \\
Ours (ResNet)     & x                & x              & 0.1694\tiny$\pm$0.0018                  & 0.5082\tiny$\pm$0.0042                  \\
Ours (FT-Trans)   & x                & x              & 0.1760{\tiny{$\pm$0.0018}}          & 0.5205{\tiny{$\pm$0.0038}}          \\ \hline
\end{tabular}

\caption{Comparison of \textbf{reliability} performance on \textbf{PLOS} prediction.}
\label{tab:model_performance_rel_plos}
\end{table}


Overall, our framework demonstrated superior performance across both tasks. In the mortality prediction task, our framework using MLP and Clinical BERT as the backbones achieved the highest F1 score (0.4629). Furthermore, with FT-Transformer and Clinical BERT, it achieved the highest BACC of 0.7672, AUROC of 0.8534, and AUPRC of 0.4977. Compared to the best baseline models, our framework improved predictive performance by approximately 1.05\% in BACC, 9.74\% in F1 score, 1.28\% in AUROC, and 6.21\% in AUPRC.
In the PLOS prediction task, our framework demonstrated exceptional performance across all evaluation metrics for predictive accuracy , achieving a BACC of 0.7027, an F1 score of 0.4019, an AUROC of 0.7743, and an AUPRC of 0.3639. Compared to the best baselines, our framework showed notable improvements: a 1.02\% increase in BACC, a 6.04\% increase in F1 score, a 0.9\% increase in AUROC, and a 2.68\% increase in AUPRC.

Over the baseline models, the multimodal approaches showed marginally worse predictive accuracy than others in both tasks.
Specifically, in the mortality prediction task, the multimodal approach with clinical BERT as the backbone achieved the highest F1 of 0.4218 and AUPRC of 0.4686. In the PLOS task, it achieved the highest AUPRC of 0.3544. 

\subsubsection{Reliability}
In clinical settings, assessing prediction reliability is as important as evaluating predictability. Table \ref{tab:model_performance_rel_mortality} and \ref{tab:model_performance_rel_plos} present the comparison of model performance based on reliability across both prediction tasks.

In the mortality prediction task, our framework achieved the lowest Brier score (0.1176) and NLL (0.3594) with MLP and Clinical BERT as the backbones.
It improved prediction reliability significantly compared to the best baseline models, reducing the Brier score by about 26.8\% and NLL by 25.0\%.
In the PLOS prediction task, our framework also showed strong performance, with a Brier score of 0.1637 and an NLL of 0.4943. These results showed notable improvements over the best baseline model, with a 15.1\% reduction in Brier score and a 13.3\% reduction in NLL.

Compared to other baselines, the multimodal approaches exhibited slightly better reliability in the mortality prediction task, achieving a Brier score of 0.1606 and an NLL of 0.4792. However, in the PLOS prediction tasks, the multimodal approaches demonstrated weaker reliability.

\begin{figure}[h]
    \centering
    \includegraphics[width=0.8\linewidth]{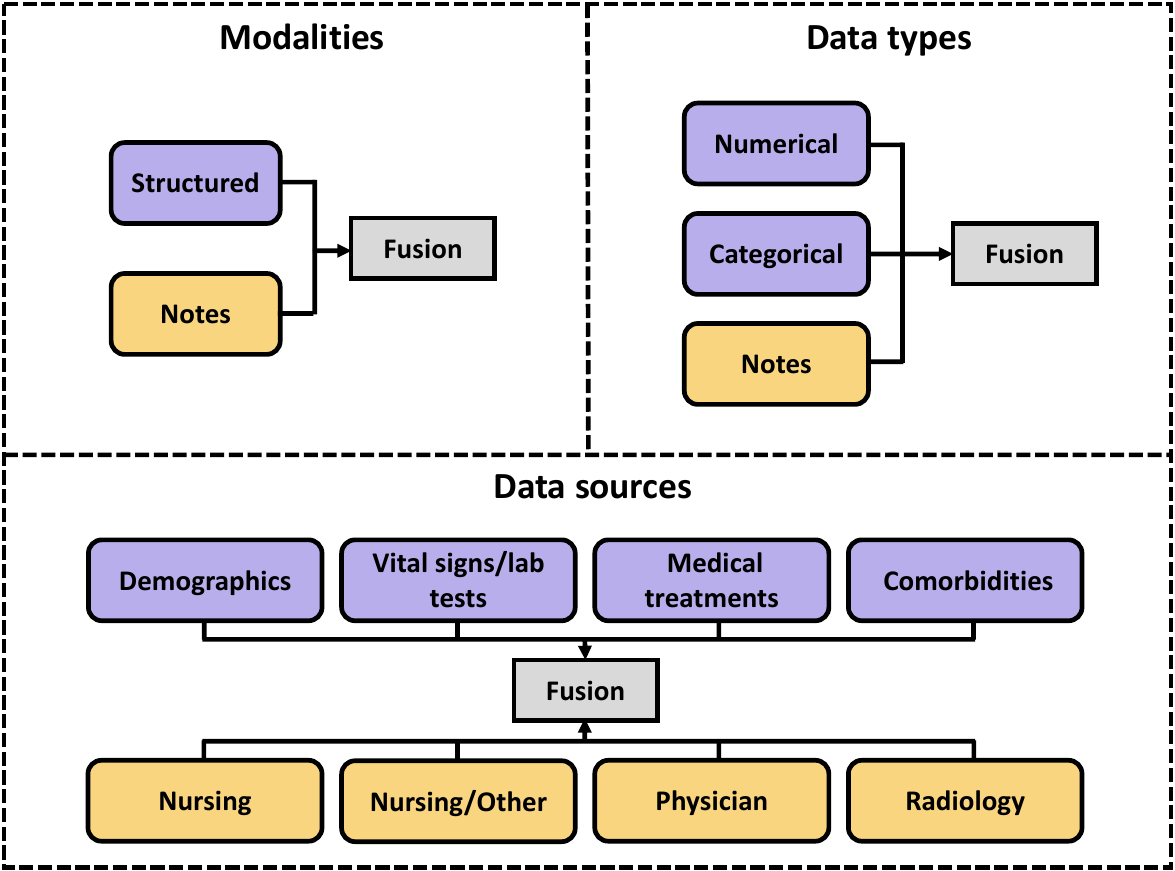}
    \caption{The illustration of different fusion settings.}
    \label{fig:fusion_settings}
\end{figure}

\subsection{Evaluation on different fusion settings}



\begin{figure}[htbp]
    \centering
    \includegraphics[width=1\linewidth]{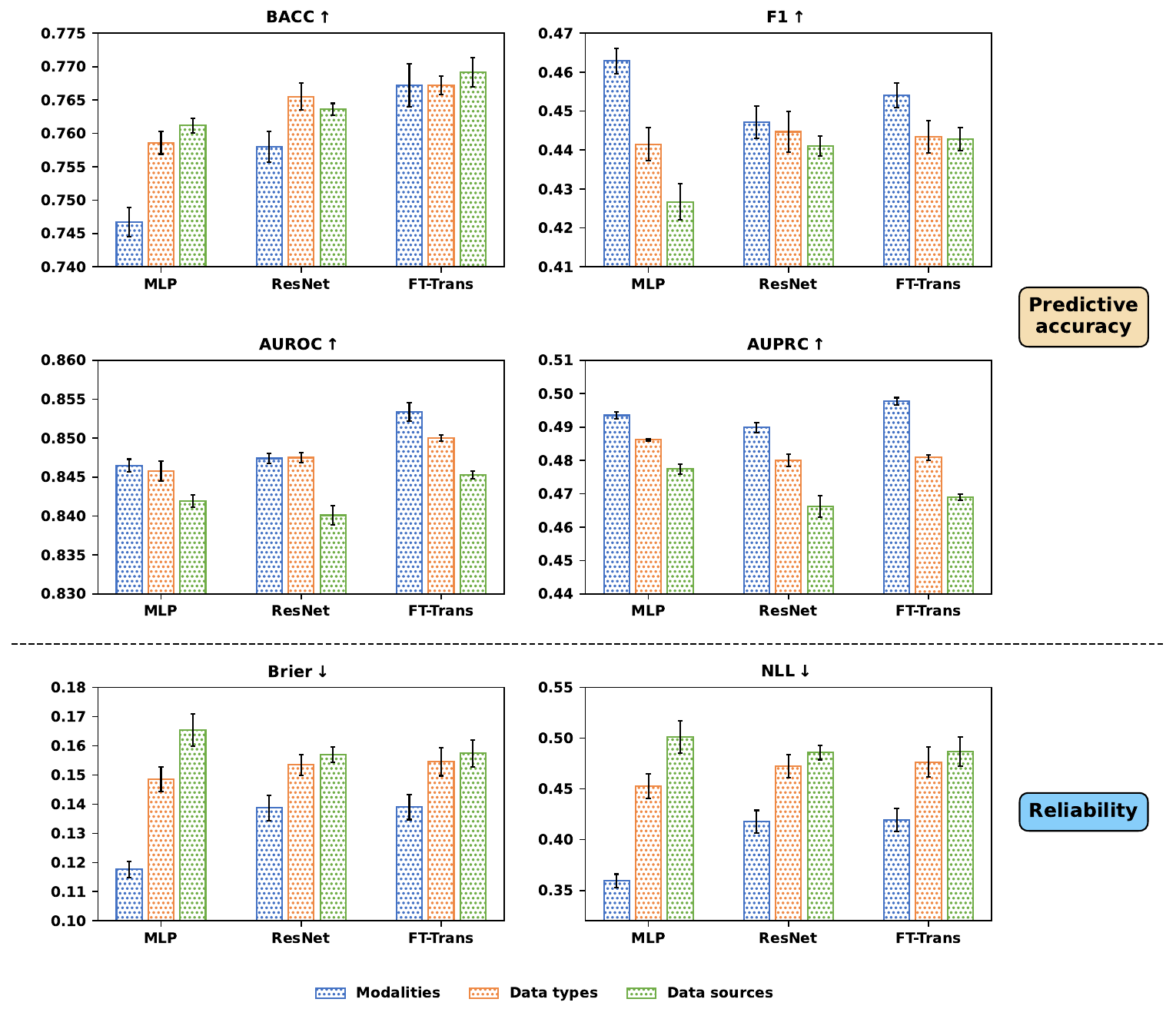}
    \caption{The evaluation of our framework using \textbf{Clinical BERT} as text encoder on different fusion settings for \textbf{mortality} prediction: (1) modalities, (2) data types, (3) data sources.}
    \label{fig:cbert_mortality_bars}
\end{figure}

\begin{figure}[htbp]
    \centering
    \includegraphics[width=1\linewidth]{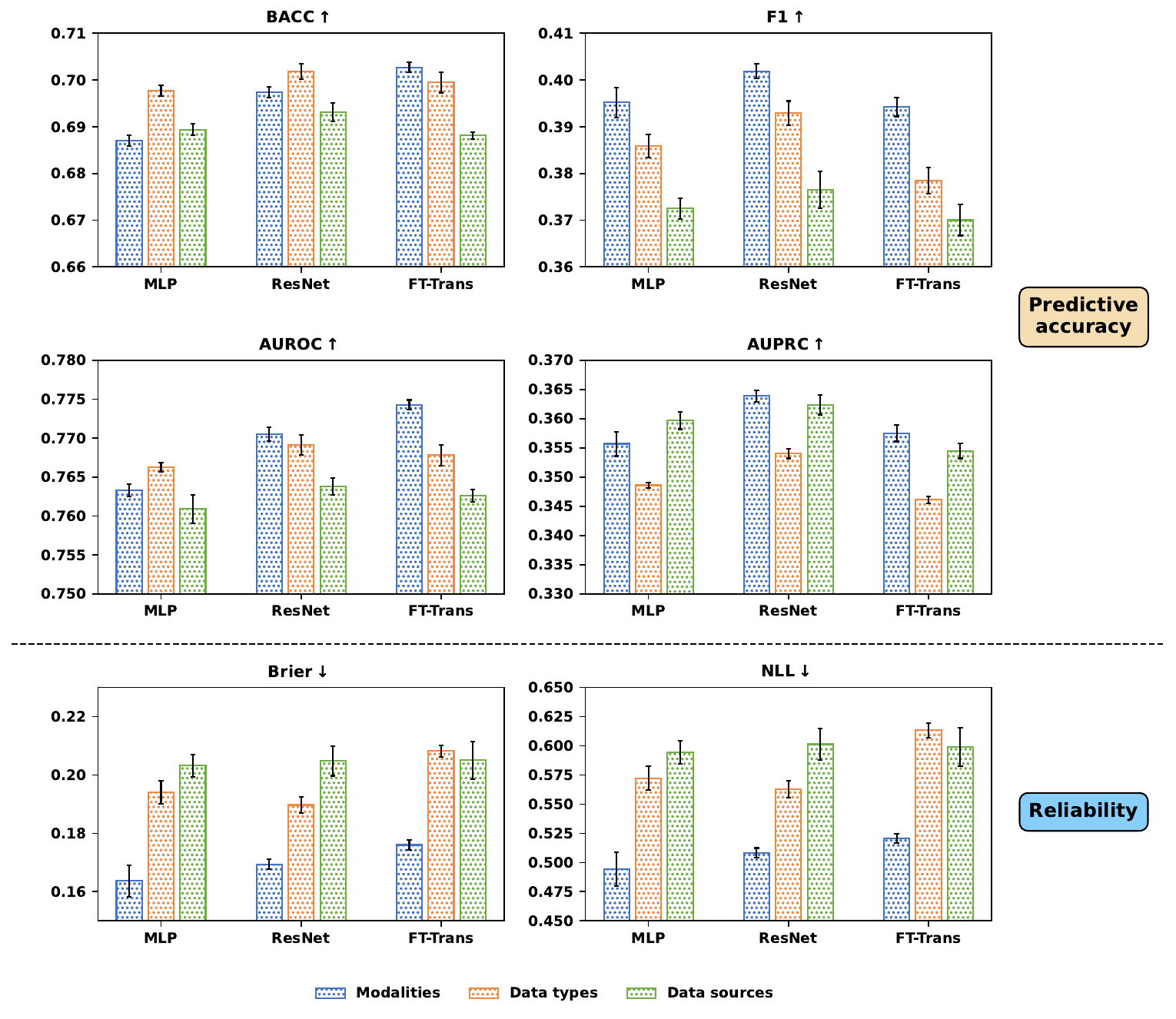}
    \caption{The evaluation of our framework using \textbf{Clinical BERT} as text encoder on different fusion settings for \textbf{PLOS} prediction: (1) modalities, (2) data types, (3) data sources.}
    \label{fig:cbert_plos_bars}
\end{figure}

We also validate the effectiveness of our fusion setting in EHR groups by comparing it with two other additional fusion settings: data types and data sources. The illustration is shown in Figure \ref{fig:fusion_settings}. For the data types fusion, we divided the structured data into two types: numerical and categorical data. For the data sources fusion, we split the structured data into four sources: demographics, vital signs/lab tests, medical treatments, and comorbidities, and categorized the clinical notes into four types: Nursing, Nursing/Other, Physician, and Radiology notes. Figures \ref{fig:cbert_mortality_bars} and \ref{fig:cbert_plos_bars} illustrate the evaluation of our framework using clinical BERT as text encoder across three fusion settings for both tasks. Evaluation results about models using BERT and BioBERT as text encoders can be found in Appendix \hyperref[appendix:different_settings]{A}.

Regarding predictive accuracy, the models achieved lower BACC across the three fusion settings for both tasks. Notably, the model based on modalities outperformed those based on data types and data sources in F1, AUROC, and AUPRC metrics. In terms of reliability evaluation metrics, the model based on modalities demonstrated superior reliability across three different structured data encoders. 
Additionally, Figure \ref{fig:cbert_mortality_bars} and \ref{fig:cbert_plos_bars} suggest a positive relationship between predictability and reliability in our framework. 

\section{DISCUSSION}

In this section, we present our findings across four key areas: the advantages of using free-text notes for mortality and PLOS prediction, the effectiveness of the evidence-based multimodal framework, the impact of various fusion settings, and the influence of encoder selection.

\subsection{Benefits of free-text notes}
From Tables \ref{tab:model_performance_pred_mortality} and \ref{tab:model_performance_pred_plos}, it is evident that integrating free-text notes with structured data enhances ICU outcome prediction, boosting predictability. Free-text notes provide information not included in structured EHR data, such as nursing details, physician documentation, and radiology reports after ICU admission. This suggests that free-text EHR notes and structured inputs can complement each other in predictive modeling, leading to improved performance.

\subsection{Effectiveness on multimodal evidence fusion}

\paragraph{Accurate and reliable ICU decision support}


The proposed framework outperforms existing multimodal approaches by leveraging belief function theory for the effective fusion of structured and unstructured EHR data. This enables more accurate and robust predictions, which is essential for clinical decision-making in ICU settings.
Notably, while the improvement in BACC is not significant (1.05\% for mortality and 1.02\% for PLOS prediction), the F1 score shows a significant increase (9.74\% for mortality and 6.04\% for PLOS prediction). This highlights the ability of the framework to identify critical ICU cases, as the F1 score prioritizes precision and recall, making it especially valuable in imbalanced datasets where positive cases are rare. Given the high-stakes nature of ICU outcomes, this improvement suggests our model enhances early identification of high-risk patients, potentially supporting timelier interventions in critical care.


Unlike traditional multimodal approaches, our evidence-based framework demonstrates greater prediction reliability, which is particularly valuable in clinical decision-making.
By effectively handling uncertainty and inconsistencies in patient data under the proposed multimodal fusion framework, our approach ensures more trustworthy predictions with lower Brier and NLL (shown in Table \ref{tab:model_performance_rel_mortality} and \ref{tab:model_performance_rel_plos}).
This enhanced reliability is crucial for ICU decision support, where inaccurate predictions can lead to overuse of critical resources or missed early interventions.

\paragraph{Efficient resource allocation}

The experimental results in Tables \ref{tab:model_performance_add_motality} and \ref{tab:model_performance_add_plos} demonstrate that while existing multimodal approaches can effectively capture true positive instances, they often come at the cost of increased false positives. In ICU settings, such false positives can lead to the unnecessary use of medical resources and equipment. In contrast, our framework achieves a better balance by demonstrating higher precision and specificity, effectively reducing false positives. This capability is crucial for ensuring that ICU resources are allocated appropriately to patients in critical need.

\begin{table}[t]


\begin{tabular}{p{2.3cm}p{0.8cm}p{0.6cm}p{1.6cm}p{1.6cm}p{1.65cm}p{1.6cm}}
\hline
\textbf{Model}    & \textbf{Struct.} & \textbf{Notes} & \textbf{Precision$\uparrow$}                           & \textbf{Recall$\uparrow$}                              & \textbf{Specificity$\uparrow$}                         & \textbf{NPV$\uparrow$}                                 \\ \hline
Random Forest     & x                &                & 0.2634{\tiny{$\pm$0.0014}}          & 0.6946{\tiny{$\pm$0.0037}}          & 0.7398{\tiny{$\pm$0.0015}}          & 0.9476{\tiny{$\pm$0.0006}}          \\
MLP               & x                &                & 0.2741{\tiny{$\pm$0.0023}}          & 0.7707{\tiny{$\pm$0.0045}}          & 0.7264{\tiny{$\pm$0.0047}}          & 0.9594{\tiny{$\pm$0.0005}}          \\
ResNet            & x                &                & \textbf{0.2809{\tiny{$\pm$0.0007}}}          & 0.7676{\tiny{$\pm$0.0036}}          & 0.7367{\tiny{$\pm$0.0017}}          & 0.9594{\tiny{$\pm$0.0005}}          \\
FT-Transformer    & x                &                & 0.2838{\tiny{$\pm$0.0015}}          & 0.7832\tiny$\pm$0.0054                  & 0.7352{\tiny{$\pm$0.0019}}          & 0.9620\tiny$\pm$0.0009                  \\ \hline
\multicolumn{7}{c}{BERT as text encoder}                                                                                                                                                                                                                                                  \\ \hline
Text encoder only &                  & x              & 0.1767{\tiny{$\pm$0.0030}}          & 0.6966{\tiny{$\pm$0.0225}}          & 0.5634{\tiny{$\pm$0.0217}}          & 0.9330{\tiny{$\pm$0.0021}}          \\
Multimodal        & x                & x              & 0.2761{\tiny{$\pm$0.0034}}          & \textbf{0.7807{\tiny{$\pm$0.0063}}}          & 0.7255{\tiny{$\pm$0.0064}}          & \textbf{0.9611{\tiny{$\pm$0.0008}}}          \\
\hdashline[2.5pt/5pt]
Ours (MLP)        & x                & x              & \textbf{0.3297{\tiny{$\pm$0.0076}}}          & 0.6957{\tiny{$\pm$0.0168}}          & \textbf{0.8094{\tiny{$\pm$0.0114}}}          & 0.9522{\tiny{$\pm$0.0020}}          \\
Ours (ResNet)     & x                & x              & 0.3179{\tiny{$\pm$0.0073}}          & 0.7206{\tiny{$\pm$0.0168}}          & 0.7916{\tiny{$\pm$0.0118}}          & 0.9550{\tiny{$\pm$0.0020}}          \\
Ours (FT-Trans)   & x                & x              & 0.3163{\tiny{$\pm$0.0080}}          & 0.7433{\tiny{$\pm$0.0109}}          & 0.7836{\tiny{$\pm$0.0110}}          & 0.9580{\tiny{$\pm$0.0012}}          \\ \hline
\multicolumn{7}{c}{BioBERT as text encoder}                                                                                                                                                                                                                                               \\ \hline
Text encoder only &                  & x              & 0.1793{\tiny{$\pm$0.0011}}          & 0.6029{\tiny{$\pm$0.0278}}          & 0.6299{\tiny{$\pm$0.0194}}          & 0.9225{\tiny{$\pm$0.0028}}          \\
Multimodal        & x                & x              & 0.2741{\tiny{$\pm$0.0031}}          & \cellcolor{mycolor} \textbf{0.7934{\tiny{$\pm$0.0065}}} & 0.7182{\tiny{$\pm$0.0067}}          & \cellcolor{mycolor} \textbf{0.9629{\tiny{$\pm$0.0008}}} \\
\hdashline[2.5pt/5pt]
Ours (MLP)        & x                & x              & \textbf{0.3377\tiny$\pm$0.0115}                  & 0.6792{\tiny{$\pm$0.0218}}          & \textbf{0.8191\tiny$\pm$0.0148}                  & 0.9503{\tiny{$\pm$0.0023}}          \\
Ours (ResNet)     & x                & x              & 0.3244{\tiny{$\pm$0.0100}}          & 0.7092{\tiny{$\pm$0.0219}}          & 0.8000{\tiny{$\pm$0.0150}}          & 0.9538{\tiny{$\pm$0.0025}}          \\
Ours (FT-Trans)   & x                & x              & 0.3282{\tiny{$\pm$0.0015}}          & 0.7193{\tiny{$\pm$0.0055}}          & 0.8027{\tiny{$\pm$0.0027}}          & 0.9553{\tiny{$\pm$0.0007}}          \\ \hline
\multicolumn{7}{c}{Clinical BERT as text encoder}                                                                                                                                                                                                                                         \\ \hline
Text encoder only &                  & x              & 0.1962{\tiny{$\pm$0.0012}}          & 0.7158{\tiny{$\pm$0.0080}}          & 0.6071{\tiny{$\pm$0.0073}}          & 0.9410{\tiny{$\pm$0.0009}}          \\
Multimodal        & x                & x              & 0.2908{\tiny{$\pm$0.0027}}          & \textbf{0.7678{\tiny{$\pm$0.0063}}}          & 0.7489{\tiny{$\pm$0.0054}}          & \textbf{0.9601{\tiny{$\pm$0.0008}}}          \\
\hdashline[2.5pt/5pt]
Ours (MLP)        & x                & x              & \cellcolor{mycolor} \textbf{0.3609{\tiny{$\pm$0.0075}}} & 0.6478{\tiny{$\pm$0.0123}}          & \cellcolor{mycolor} \textbf{0.8454{\tiny{$\pm$0.0079}}} & 0.9472{\tiny{$\pm$0.0013}}          \\
Ours (ResNet)     & x                & x              &  0.3260{\tiny{$\pm$0.0067}}           & 0.7151{\tiny{$\pm$0.0131}}          & 0.8010{\tiny{$\pm$0.0097}}          & 0.9546{\tiny{$\pm$0.0015}}          \\
Ours (FT-Trans)   & x                & x              & 0.3289{\tiny{$\pm$0.0070}}          & 0.7371{\tiny{$\pm$0.0171}}          & 0.7974{\tiny{$\pm$0.0108}}          & 0.9578{\tiny{$\pm$0.0021}}          \\ \hline
\end{tabular}

\caption{Comparison of class-specific prediction accuracy on \textbf{mortality} prediction.}
\label{tab:model_performance_add_motality}
\end{table}

\begin{table}[t]


\begin{tabular}{p{2.3cm}p{0.8cm}p{0.6cm}p{1.6cm}p{1.6cm}p{1.65cm}p{1.6cm}}
\hline
\textbf{Model}    & \textbf{Struct.} & \textbf{Notes} & \textbf{Precision$\uparrow$}                           & \textbf{Recall$\uparrow$}                              & \textbf{Specificity$\uparrow$}                         & \textbf{NPV$\uparrow$}                                 \\ \hline
Random Forest     & x                &                & 0.2360{\tiny{$\pm$0.0045}}          & 0.6744{\tiny{$\pm$0.0131}}          & 0.6616{\tiny{$\pm$0.0150}}          & 0.9295{\tiny{$\pm$0.0014}}          \\
MLP               & x                &                & 0.2615{\tiny{$\pm$0.0013}}          & 0.6542{\tiny{$\pm$0.0034}}          & 0.7147{\tiny{$\pm$0.0030}}          & 0.9305{\tiny{$\pm$0.0004}}          \\
ResNet            & x                &                & 0.2673{\tiny{$\pm$0.0016}}          & 0.6420{\tiny{$\pm$0.0039}}          & 0.7282{\tiny{$\pm$0.0025}}          & 0.9295{\tiny{$\pm$0.0006}}          \\
FT-Transformer    & x                &                & 0.2603{\tiny{$\pm$0.0026}}          & 0.6974{\tiny{$\pm$0.0061}}          & 0.6938{\tiny{$\pm$0.0067}}          & 0.9369{\tiny{$\pm$0.0007}}          \\ \hline
\multicolumn{7}{c}{BERT as text encoder}                                                                                                                                                                                                                                                  \\ \hline
Text encoder only &                  & x              & 0.2004{\tiny{$\pm$0.0031}}          & 0.5685{\tiny{$\pm$0.0147}}          & 0.6488{\tiny{$\pm$0.0160}}          & 0.9070{\tiny{$\pm$0.0009}}          \\
Multimodal        & x                & x              & 0.2568{\tiny{$\pm$0.0029}}          & 0.6892{\tiny{$\pm$0.0141}}          & 0.6915{\tiny{$\pm$0.0108}}          & 0.9347{\tiny{$\pm$0.0022}}          \\
\hdashline[2.5pt/5pt]
Ours (MLP)        & x                & x              & \textbf{0.2809{\tiny{$\pm$0.0038}}}          & 0.6218{\tiny{$\pm$0.0050}}          & \textbf{0.7538{\tiny{$\pm$0.0064}}}          & 0.9281{\tiny{$\pm$0.0005}}          \\
Ours (ResNet)     & x                & x              & 0.2787{\tiny{$\pm$0.0071}}          & 0.6460{\tiny{$\pm$0.0158}}          & 0.7402{\tiny{$\pm$0.0148}}          & 0.9314{\tiny{$\pm$0.0016}}          \\
Ours (FT-Trans)   & x                & x              & 0.2595{\tiny{$\pm$0.0036}}          & \textbf{0.7180\tiny$\pm$0.0116}                  & 0.6830{\tiny{$\pm$0.0110}}          & \cellcolor{mycolor} \textbf{0.9402{\tiny{$\pm$0.0014}}} \\ \hline
\multicolumn{7}{c}{BioBERT as text encoder}                                                                                                                                                                                                                                               \\ \hline
Text encoder only &                  & x              & 0.2006{\tiny{$\pm$0.0022}}          & 0.5504{\tiny{$\pm$0.0160}}          & 0.6610{\tiny{$\pm$0.0141}}          & 0.9051{\tiny{$\pm$0.0013}}          \\
Multimodal        & x                & x              & 0.2554{\tiny{$\pm$0.0034}}          & \textbf{0.6943{\tiny{$\pm$0.0136}}}          & 0.6868{\tiny{$\pm$0.0114}}          & 0.9358{\tiny{$\pm$0.0018}}          \\
\hdashline[2.5pt/5pt]
Ours (MLP)        & x                & x              & \textbf{0.2892{\tiny{$\pm$0.0048}}}          & 0.6022{\tiny{$\pm$0.0061}}          & \textbf{0.7710{\tiny{$\pm$0.0071}}}          & 0.9262{\tiny{$\pm$0.0007}}          \\
Ours (ResNet)     & x                & x              & 0.2848{\tiny{$\pm$0.0059}}          & 0.6212{\tiny{$\pm$0.0126}}          & 0.7580{\tiny{$\pm$0.0119}}          & 0.9285{\tiny{$\pm$0.0012}}          \\
Ours (FT-Trans)   & x                & x              & 0.2668{\tiny{$\pm$0.0042}}          & 0.6914{\tiny{$\pm$0.0106}}          & 0.7059{\tiny{$\pm$0.0109}}          & \textbf{0.9368{\tiny{$\pm$0.0012}}}          \\ \hline
\multicolumn{7}{c}{Clinical BERT as text encoder}                                                                                                                                                                                                                                         \\ \hline
Text encoder only &                  & x              & 0.2208{\tiny{$\pm$0.0024}}          & 0.6251{\tiny{$\pm$0.0133}}          & 0.6587{\tiny{$\pm$0.0116}}          & 0.9193{\tiny{$\pm$0.0014}}          \\
Multimodal        & x                & x              & 0.2507{\tiny{$\pm$0.0045}}          & \cellcolor{mycolor} \textbf{0.7201{\tiny{$\pm$0.0188}}} & 0.6664{\tiny{$\pm$0.0163}}          & \textbf{0.9394\tiny$\pm$0.0024}                  \\
\hdashline[2.5pt/5pt]
Ours (MLP)        & x                & x              & \cellcolor{mycolor} \textbf{0.2980{\tiny{$\pm$0.0075}}} & 0.5899{\tiny{$\pm$0.0134}}          & \cellcolor{mycolor} \textbf{0.7841{\tiny{$\pm$0.0117}}} & 0.9254{\tiny{$\pm$0.0012}}          \\
Ours (ResNet)     & x                & x              & 0.2968\tiny$\pm$0.0029                  & 0.6227{\tiny{$\pm$0.0073}}          & 0.7720\tiny$\pm$0.0058                  & 0.9299{\tiny{$\pm$0.0008}}          \\
Ours (FT-Trans)   & x                & x              & 0.2782{\tiny{$\pm$0.0026}}          & 0.6766{\tiny{$\pm$0.0046}}          & 0.7288{\tiny{$\pm$0.0050}}          & 0.9359{\tiny{$\pm$0.0005}}          \\ \hline
\end{tabular}

\caption{Comparison of class-specific prediction accuracy on \textbf{PLOS} prediction.}
\label{tab:model_performance_add_plos}
\end{table}

\subsection{Analysis on different fusion settings}
\begin{figure}[htbp]
    \centering
    \includegraphics[width=0.95\linewidth]{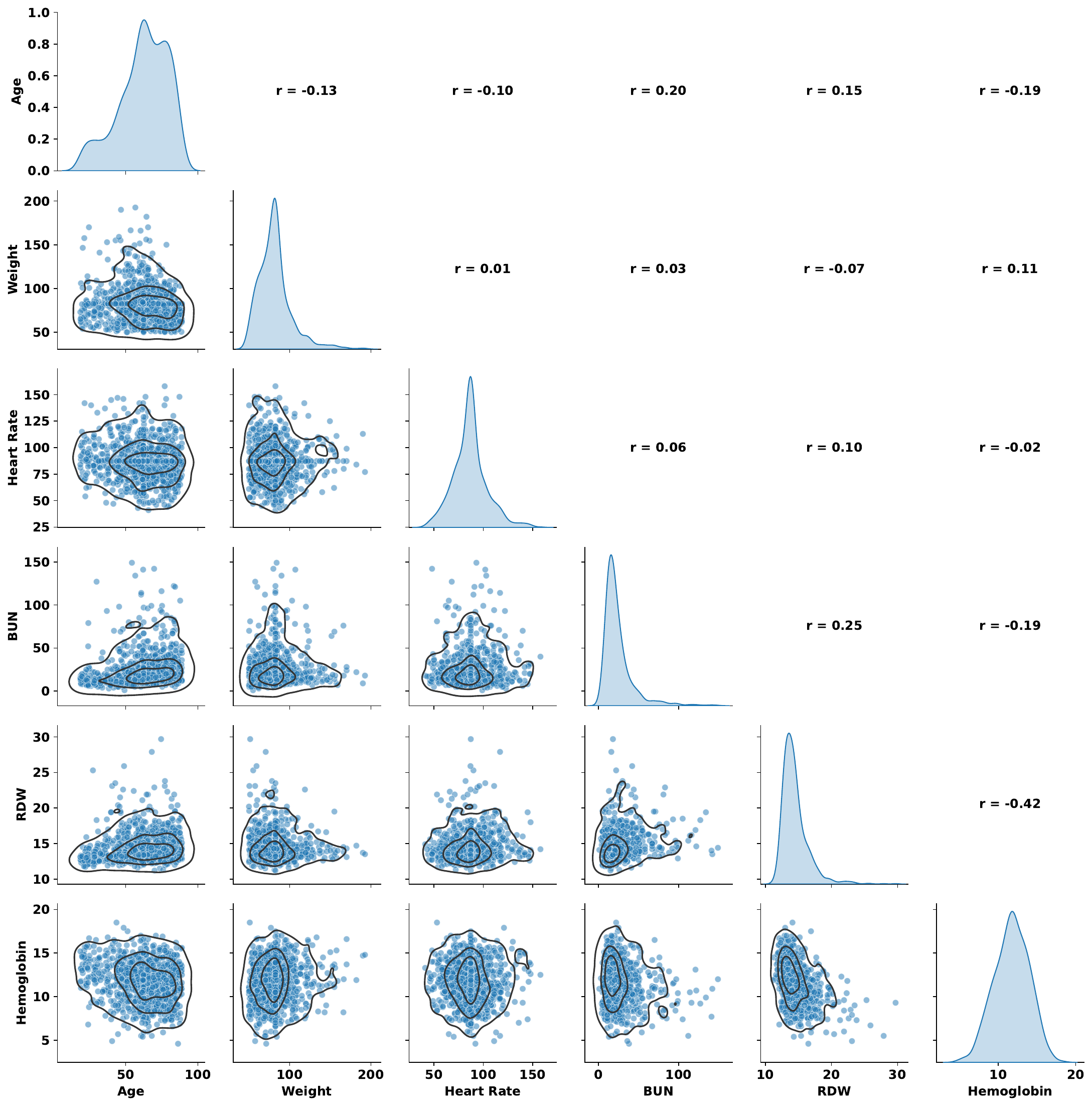}
    \caption{The pair plots and correlation coefficients of some structured features from different sources.}
    \label{fig:structured_indep}
\end{figure}

\begin{figure}[htbp]
    \centering
    \includegraphics[width=1\linewidth]{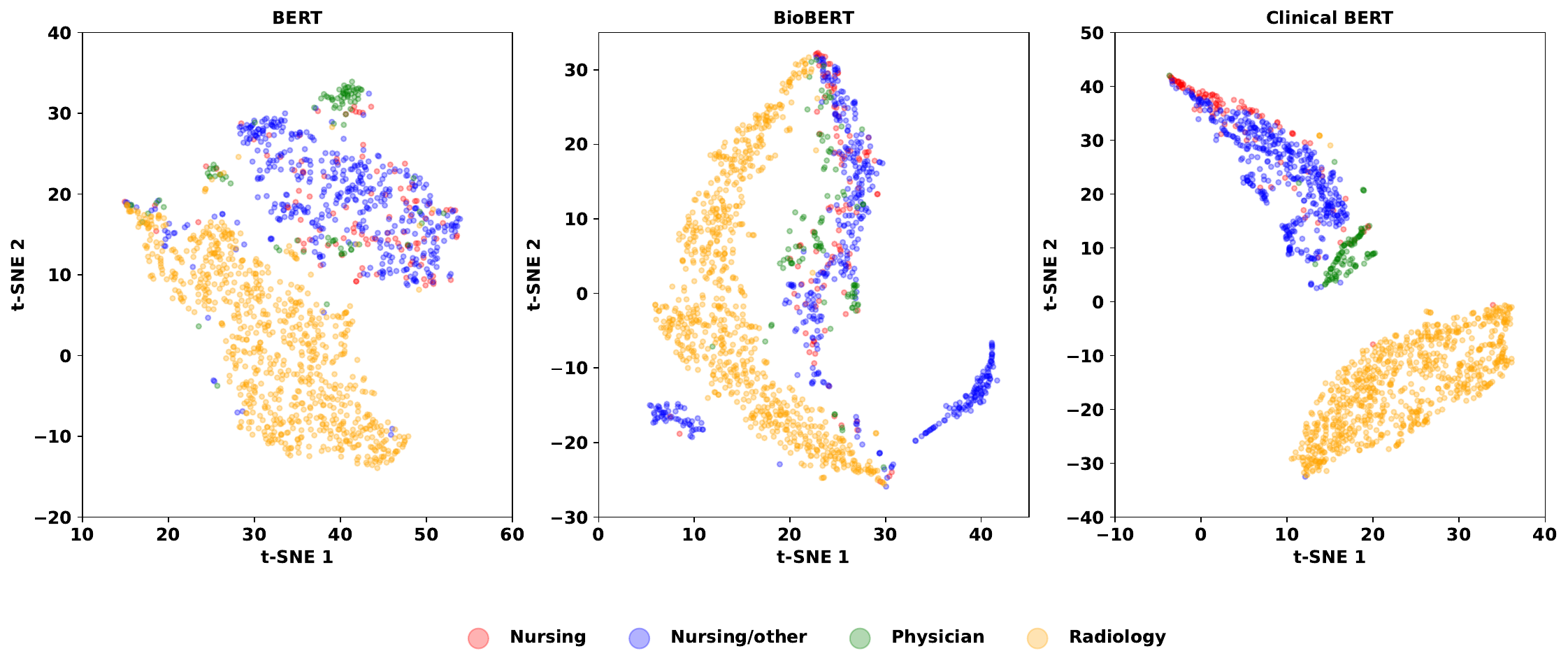}
    \caption{The t-SNE visualization on extracted embeddings  of four different types of free-texts in EHRs: Nursing, Nursing/other, Physician, Radiology.}
    \label{fig:text_indep}
\end{figure}

To explore the performance of our framework across different fusion settings, it is evident that models based on modalities achieved higher F1 scores but lower BACCs compared to the other two fusion settings. This discrepancy arises from the metrics' focus: the F1 score emphasizes performance on the positive class by balancing precision and recall, while BACC provides an overall assessment of recall across all classes. Thus, models based on modalities are more effective at identifying ICU outcomes, likely due to the enhanced integration of information from independent sources facilitated by belief function theory.
This observation aligns with our analysis of data independence. Figures \ref{fig:structured_indep} and \ref{fig:text_indep} visualize the independence of structured features and four types of free-text notes, respectively. In Figure \ref{fig:structured_indep}, correlation coefficients confirm that features within structured data are not independent. Meanwhile, Figure \ref{fig:text_indep} shows that Radiology notes form a distinct cluster, while the other types of notes overlap, indicating a lack of independence among them. 
Moreover, the difference in BACC performance using the FT-Transformer across three fusion settings is smaller than with the MLP. This is likely due to the stronger predictive capability of the FT-Transformer.



\subsection{Influence of encoders selection}
The evaluations presented in Table \ref{tab:model_performance_pred_mortality} and \ref{tab:model_performance_pred_plos} reveal that the choice of encoders plays a critical role in determining predictability. Among the assessed models, the FT-Transformer stands out as highly effective for structured tabular data, aligning with the findings of \cite{gorishniy2021revisiting}. This effectiveness can be attributed to the transformer's capability to capture complex relationships among transformed numerical and categorical features, which enhances its predictive power.

For pre-trained language encoders, Clinical BERT demonstrates the best performance in extracting clinical information from free-text notes. Its superiority stems from its unique pre-training on clinical text from the MIMIC-III database. This specialized pre-training enables Clinical BERT to generate more informative embeddings by leveraging its pre-learned clinical knowledge and domain-specific term embeddings, resulting in improved model performance.

\section{CONCLUSION}

In this paper, we address the challenge of accurately and reliably predicting ICU outcomes by introducing a multimodal framework based on belief function theory that models both structured EHR data and free-text EHR notes. Our framework transforms deep features extracted from these two modalities into evidence through the evidence mapping module, which is then fused using Dempster's rule to make final predictions. Through experiments on the MIMIC-III dataset, we demonstrate the effectiveness of our framework in terms of predictive accuracy and reliability. The study highlights its capability in managing heterogeneous multimodal EHR data, reducing false positives and potentially improving the allocation of medical resources in the ICU.

While this paper focuses on binary classification tasks, many clinical applications require solutions for multiclass tasks (e.g., disease diagnosis) and continuous regression tasks (e.g., survival prediction). These are equally important and relevant for advancing clinical practice. In the future, we plan to expand our framework by incorporating additional data modalities, such as time series and medical images, to provide deeper clinical insights and enhance model performance. We also aim to extend the framework to handle multimodal EHR multiclass tasks, offering valuable predictive guidance for complex clinical scenarios. Additionally, we intend to investigate regression tasks, leveraging the recently introduced Epistemic Random Fuzzy Set (ERFS) theory \cite{denoeux2021belief, denoeux2023reasoning} and further building on developments in evidential regression \cite{huang2024evidential}.

\backmatter

\bmhead{Acknowledgements}

This research is supported by the National Research Foundation, Singapore under its AI Singapore Programme (AISG Award No: AISG-GC-2019-001-2B).

\bmhead{Code Availability}
The code implementation is available on Github repository (\href{https://github.com/yuchengruan/evid_multimodal_ehr}{https://github.com/yuchengruan/evid\textunderscore multimodal\textunderscore ehr}).

\section*{Declarations}

\bmhead{Ethics Approval}
Not applicable.
\bmhead{Conflict of Interest}
The authors declare no competing interests.


\bibliography{sn-bibliography}

\newpage

\appendix
\section{Evaluation on different fusion settings using BERT and BioBERT as text encoders \label{appendix:different_settings}}
\begin{figure}[htbp]
    \centering
    \includegraphics[width=1\linewidth]{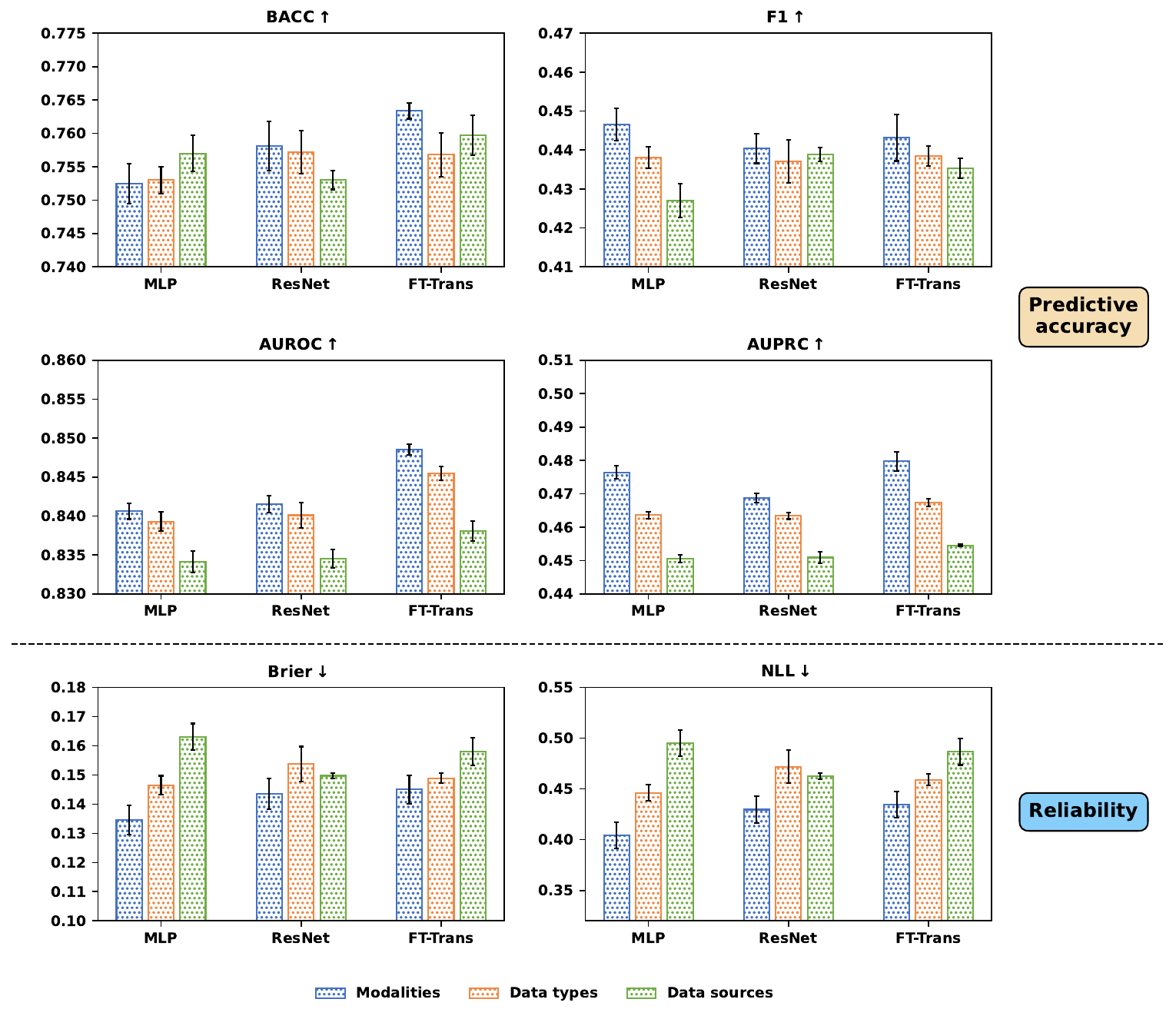}
    \caption{The evaluation of our framework using \textbf{BERT} as text encoder on different fusion settings for \textbf{mortality} prediction: (1) modalities, (2) data types, (3) data sources.}
    \label{fig:bert_mortality_bars}
\end{figure}

\begin{figure}[htbp]
    \centering
    \includegraphics[width=1\linewidth]{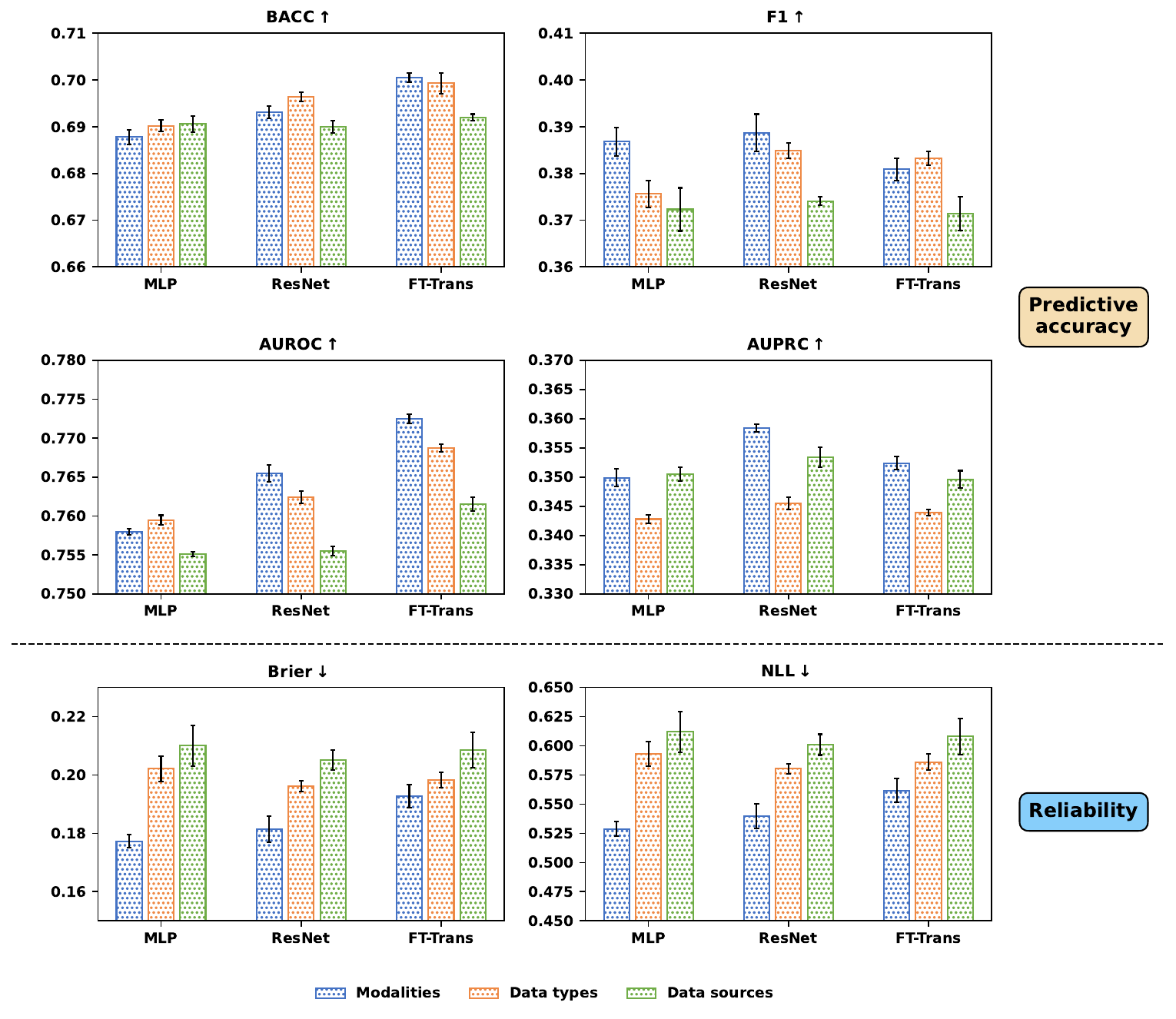}
    \caption{The evaluation of our framework using \textbf{BERT} as text encoder on different fusion settings for \textbf{PLOS} prediction: (1) modalities, (2) data types, (3) data sources.}
    \label{fig:bert_plos_bars}
\end{figure}

\begin{figure}[htbp]
    \centering
    \includegraphics[width=1\linewidth]{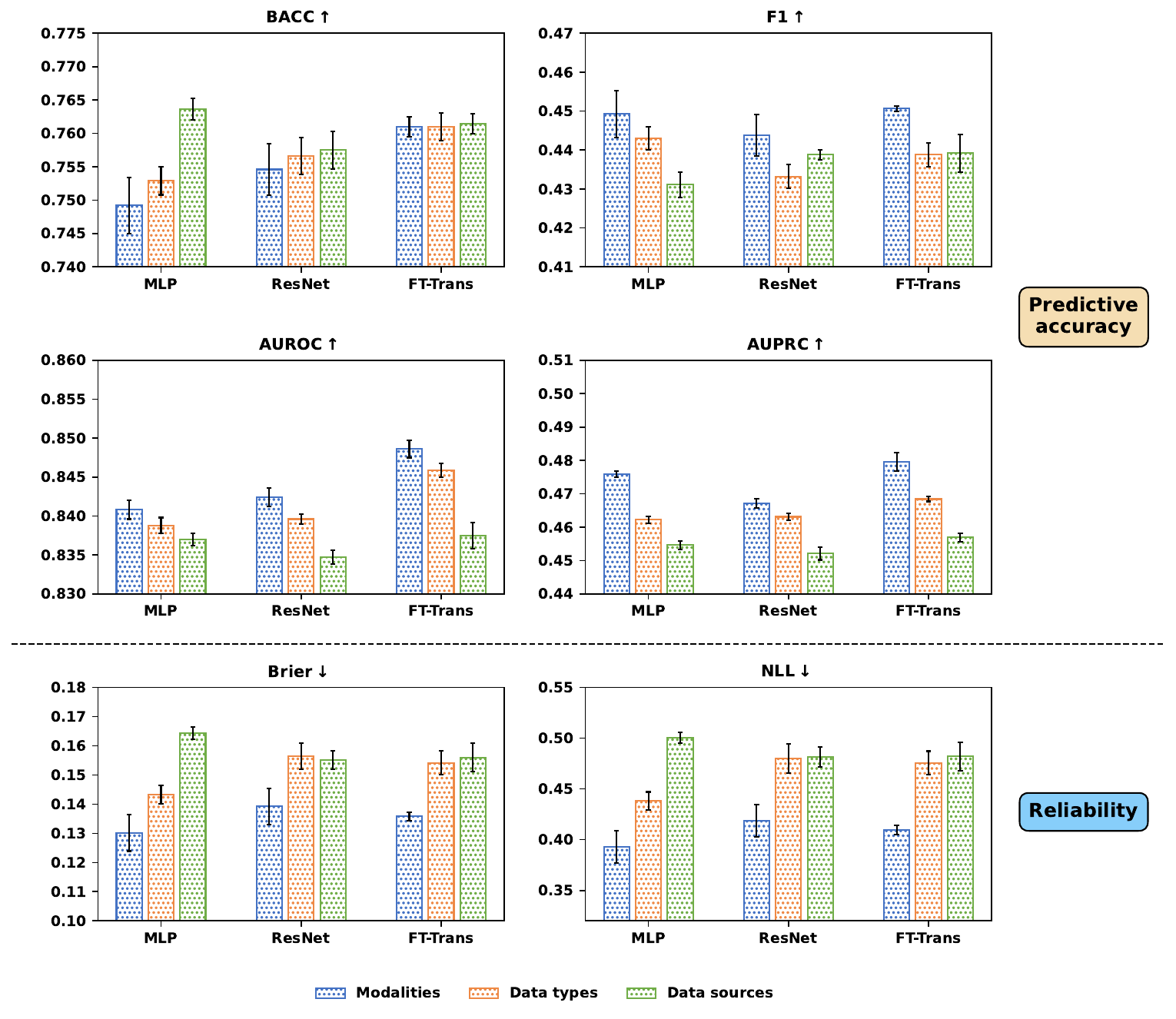}
    \caption{The evaluation of our framework using \textbf{BioBERT} as text encoder on different fusion settings for \textbf{mortality} prediction: (1) modalities, (2) data types, (3) data sources.}
    \label{fig:biobert_mortality_bars}
\end{figure}

\begin{figure}[htbp]
    \centering
    \includegraphics[width=1\linewidth]{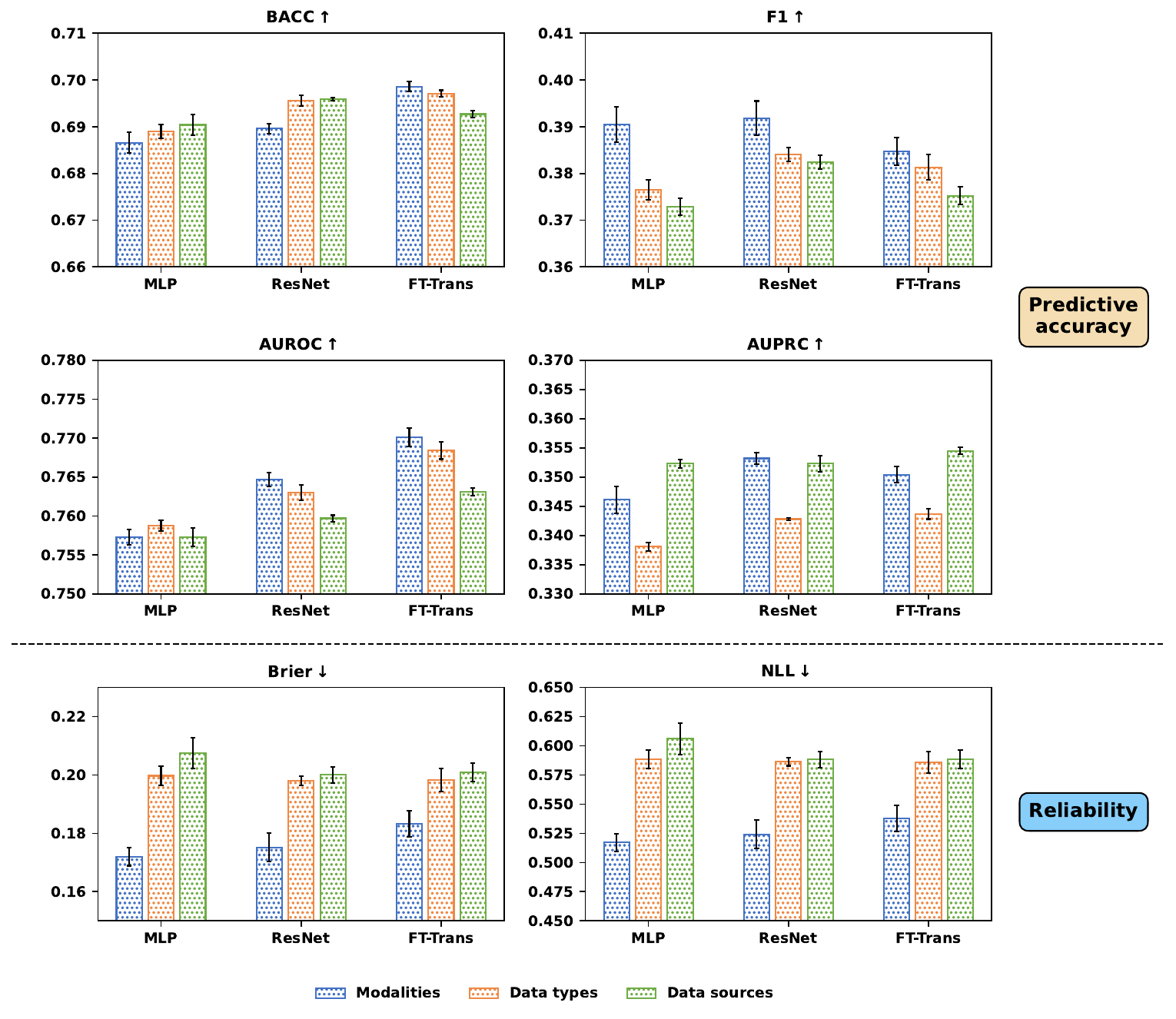}
    \caption{The evaluation of our framework using \textbf{BioBERT} as text encoder on different fusion settings for \textbf{PLOS} prediction: (1) modalities, (2) data types, (3) data sources.}
    \label{fig:biobert_plos_bars}
\end{figure}

\end{document}